%% file: main.tex
\documentclass[twocolumn]{aastex631}

\usepackage[T1]{fontenc}
\usepackage{lmodern}

\usepackage{xspace}
\usepackage{amsmath}
\usepackage{amssymb}
\usepackage{url}
\usepackage{graphicx}
\usepackage{hyperref}

\newcommand{\dphn}{\phn\phn}
\newcommand{\tphn}{\dphn\phn}
\newcommand{\phne}{\phm{$(1)$}}
\newcommand{\phnm}{\phm{$-$}}

\newcommand{\tempotwo}{\textsc{tempo2}\xspace}
\newcommand{\psrchive}{\textsc{psrchive}\xspace}

\newcommand{\psra}{J0032+6946\xspace}
\newcommand{\psrb}{J0141+6303\xspace}
\newcommand{\psrc}{J0214+5222\xspace}

\newcommand{\psre}{J0636+5128\xspace}
\newcommand{\psrf}{J0957$-$0619\xspace}
\newcommand{\psrg}{J1239+3239\xspace}
\newcommand{\psrh}{J1327+3423\xspace}
\newcommand{\psri}{J1434+7257\xspace}

\newcommand{\psrk}{J1530$-$2114\xspace}
\newcommand{\psrl}{J1816+4510\xspace}
\newcommand{\psrm}{J1913+3732\xspace}

\newcommand{\psrr}{J2145+2158\xspace}
\newcommand{\psrs}{J2210+5712\xspace}

\newcommand{\PSRa}{PSR~\psra}
\newcommand{\PSRb}{PSR~\psrb}
\newcommand{\PSRc}{PSR~\psrc}

\newcommand{\PSRe}{PSR~\psre}
\newcommand{\PSRf}{PSR~\psrf}
\newcommand{\PSRg}{PSR~\psrg}
\newcommand{\PSRh}{PSR~\psrh}
\newcommand{\PSRi}{PSR~\psri}

\newcommand{\PSRk}{PSR~\psrk}
\newcommand{\PSRl}{PSR~\psrl}
\newcommand{\PSRm}{PSR~\psrm}

\newcommand{\PSRr}{PSR~\psrr}
\newcommand{\PSRs}{PSR~\psrs}

\newcommand{\dmu}{\textrm{pc\,cm}$^{-3}$\xspace}
\newcommand{\ergpersec}{\textrm{erg\,s}$^{-1}$\xspace}
\newcommand{\pmu}{\textrm{mas\,yr}$^{-1}$\xspace}
\newcommand{\Msun}{\textrm{M}$_\sun$\xspace}
\newcommand{\Rsun}{\textrm{R}$_\sun$\xspace}
\newcommand{\us}{$\mu$\textrm{s}\xspace}
\newcommand{\nudot}{$\dot{\nu}$\xspace}
\newcommand{\pdot}{$\dot{P}$\xspace}
\newcommand{\ps}{$\dot{P}_\mathrm{S}$\xspace}
\newcommand{\pg}{$\dot{P}_\mathrm{G}$\xspace}
\newcommand{\pint}{$\dot{P}_\mathrm{int}$\xspace}

\newcommand{\km}{\textrm{km}}

\newcommand{\kpc}{\textrm{kpc}}

\newcommand{\MHz}{\textrm{MHz}}

\newcommand{\ms}{\textrm{ms}}
\newcommand{\s}{\textrm{s}}

\newcommand{\yr}{\textrm{yr}}

\newcommand{\erg}{\textrm{erg}}
\newcommand{\gauss}{\textrm{G}}

\newcommand{\dre}{$\Delta_\mathrm{R}$\xspace}

\shorttitle{21 New Pulsar Timing Solutions} 
\shortauthors{Fiore et al.}

\begin{document}

\title{The Green Bank North Celestial Cap Survey. VIII. 21 New Pulsar Timing Solutions}

\correspondingauthor{W.~Fiore}
\email{wcf0002@mix.wvu.edu}

\author[0000-0001-5645-5336]{W.~Fiore}
\affiliation{Dept.~of Physics and Astronomy, West Virginia University, P.O.~Box 6315, Morgantown, WV 26506, USA}
\affiliation{Center for Gravitational Waves and Cosmology, West Virginia University, Chestnut Ridge Research Building,\\ Morgantown, WV 26506, USA}

\author[0000-0002-2034-2986]{L.~Levin}
\affiliation{Jodrell Bank Centre for Astrophysics, School of Physics and Astronomy, The University of Manchester, Manchester, M13 9PL, UK}

\author[0000-0001-7697-7422]{M.~A.~McLaughlin}
\affiliation{Dept.~of Physics and Astronomy, West Virginia University, P.O.~Box 6315, Morgantown, WV 26506, USA}
\affiliation{Center for Gravitational Waves and Cosmology, West Virginia University, Chestnut Ridge Research Building,\\ Morgantown, WV 26506, USA}

\author[0000-0002-8935-9882]{A.~Anumarlapudi}
\affiliation{Center for Gravitation, Cosmology, and Astrophysics, Dept.~of Physics, University of Wisconsin-Milwaukee,\\ P.O.~Box 413, Milwaukee, WI 53201, USA}

\author[0000-0001-6295-2881]{D.~L.~Kaplan}
\affiliation{Center for Gravitation, Cosmology, and Astrophysics, Dept.~of Physics, University of Wisconsin-Milwaukee,\\ P.O.~Box 413, Milwaukee, WI 53201, USA}

\author[0000-0002-1075-3837]{J.~K.~Swiggum}
\affiliation{Center for Gravitation, Cosmology, and Astrophysics, Dept.~of Physics, University of Wisconsin-Milwaukee,\\ P.O.~Box 413, Milwaukee, WI 53201, USA}
\affiliation{Dept.~of Physics, 730 High St., Lafayette College, Easton, PA 18042, USA}

\author[0000-0001-5134-3925]{G.~Y.~Agazie}
\affiliation{Center for Gravitation, Cosmology, and Astrophysics, Dept.~of Physics, University of Wisconsin-Milwaukee,\\ P.O.~Box 413, Milwaukee, WI 53201, USA}

\author{R.~Bavisotto}
\affiliation{Dept.~of Chemistry and Biochemistry \& Laboratory for Surface Studies, University of Wisconsin-Milwaukee, Milwaukee, WI 53211, USA}

\author[0000-0002-3426-7606]{P.~Chawla}
\affiliation{Dept.~of Physics and McGill Space Institute, McGill Univ., Montreal, QC H3A 2T8, Canada}

\author[0000-0002-2185-1790]{M.~E.~DeCesar}
\affiliation{George Mason University, VA 22030, resident at the U.S.~Naval Research Laboratory, Washington, D.C.~20375, USA}

\author[0000-0001-8885-6388]{T.~Dolch}
\affiliation{Dept.~of Physics, Hillsdale College, 33 E.~College Street, Hillsdale, MI 49242, USA}
\affiliation{Eureka Scientific, 2452 Delmer St., Suite 100, Oakland, CA 94602-3017, USA}

\author[0000-0001-8384-5049]{E.~Fonseca}
\affiliation{Dept.~of Physics and Astronomy, West Virginia University, P.O.~Box 6315, Morgantown, WV 26506, USA}
\affiliation{Center for Gravitational Waves and Cosmology, West Virginia University, Chestnut Ridge Research Building,\\ Morgantown, WV 26506, USA}

\author[0000-0001-9345-0307]{V.~M.~Kaspi}
\affiliation{Dept.~of Physics and McGill Space Institute, McGill Univ., Montreal, QC H3A 2T8, Canada}

\author{Z.~Komassa}
\altaffiliation{Milwaukee Tool, 13135 W.~Lisbon Road, Brookfield WI 53005, USA}
\affiliation{Center for Gravitation, Cosmology, and Astrophysics, Dept.~of Physics, University of Wisconsin-Milwaukee,\\ P.O.~Box 413, Milwaukee, WI 53201, USA}

\author[0000-0001-8864-7471]{V.~I.~Kondratiev}
\affiliation{ASTRON, the Netherlands Institute for Radio Astronomy, Oude Hoogeveensedijk 4, 7991 PD Dwingeloo, The Netherlands}

\author[0000-0001-8503-6958]{J.~van Leeuwen}
\affiliation{ASTRON, the Netherlands Institute for Radio Astronomy, Oude Hoogeveensedijk 4, 7991 PD Dwingeloo, The Netherlands}

\author[0000-0002-2972-522X]{E.~F.~Lewis}
\affiliation{Dept.~of Physics and Astronomy, West Virginia University, P.O.~Box 6315, Morgantown, WV 26506, USA}
\affiliation{Center for Gravitational Waves and Cosmology, West Virginia University, Chestnut Ridge Research Building,\\ Morgantown, WV 26506, USA}

\author[0000-0001-5229-7430]{R.~S.~Lynch}
\affiliation{Green Bank Observatory, P.O.~Box 2, Green Bank, WV 24494, USA}

\author[0000-0001-5481-7559]{A.~E.~McEwen}
\affiliation{Center for Gravitation, Cosmology, and Astrophysics, Dept.~of Physics, University of Wisconsin-Milwaukee,\\ P.O.~Box 413, Milwaukee, WI 53201, USA}

\author{R.~Mundorf}
\altaffiliation{SPECS-TII Inc., 20 Cabot Blvd.~Suite \#300, Mansfield, MA 02048, USA}
\affiliation{Center for Gravitation, Cosmology, and Astrophysics, Dept.~of Physics, University of Wisconsin-Milwaukee,\\ P.O.~Box 413, Milwaukee, WI 53201, USA}
\affiliation{Dept.~of Physics and Astronomy, West Virginia University, P.O.~Box 6315, Morgantown, WV 26506, USA}

\author[0000-0002-4187-4981]{H.~Al Noori}
\affiliation{Dept.~of Physics, University of California, Santa Barbara, CA 93106, USA}

\author[0000-0002-0430-6504]{E.~Parent}
\affiliation{Institute of Space Sciences (ICE, CSIC), Campus UAB, Carrer de Can Magrans s/n, 08193, Barcelona, Spain}
\affiliation{Institut d'Estudis Espacials de Catalunya (IEEC), Carrer Gran Capit\`a 2--4, 08034 Barcelona, Spain} 

\author[0000-0002-4795-697X]{Z.~Pleunis}
\affiliation{Dunlap Institute for Astronomy \& Astrophysics, University of Toronto, 50 St.~George Street, Toronto, Ontario, M5S 3H4, Canada}

\author[0000-0001-5799-9714]{S.~M.~Ransom}
\affiliation{National Radio Astronomy Observatory, 520 Edgemont Rd., Charlottesville, VA 22903, USA}

\author[0000-0002-7778-2990]{X.~Siemens}
\affiliation{Dept.~of Physics, Oregon State University, Corvallis, OR 97331, USA}

\author[0000-0002-6730-3298]{R.~Spiewak}
\affiliation{Jodrell Bank Centre for Astrophysics, School of Physics and Astronomy, The University of Manchester, Manchester, M13 9PL, UK}

\author[0000-0001-9784-8670]{I.~H.~Stairs}
\affiliation{Dept.~of Physics and Astronomy, University of British Columbia, 6224 Agricultural Road, Vancouver, BC V6T 1Z1, Canada}

\author[0000-0002-9507-6985]{M.~Surnis}
\affiliation{Dept.~of Physics, IISER Bhopal, Bhauri Bypass Road, Bhopal 462066, India}

\author{T.~J.~Tobin}
\affiliation{Center for Gravitation, Cosmology, and Astrophysics, Dept.~of Physics, University of Wisconsin-Milwaukee,\\ P.O.~Box 413, Milwaukee, WI 53201, USA}

\begin{abstract}
We present timing solutions for 21 pulsars discovered in 350\,MHz surveys using the Green Bank Telescope (GBT). 
All were discovered in the Green Bank North Celestial Cap pulsar survey, with the exception of \PSRf, which was found in the GBT 350\,MHz Drift-scan pulsar survey. 
The majority of our timing observations were made with the GBT at 820\,MHz.
With a spin period of 37\,ms and a 528-day orbit, \PSRa joins a small group of five other mildly recycled wide binary pulsars, for which the duration of recycling through accretion is limited by the length of the companion's giant phase.
PSRs \psrb and \psrh are new disrupted recycled pulsars.
We incorporate Arecibo observations from the NANOGrav pulsar timing array into our analysis of the latter.
We also observed \PSRh with the Long Wavelength Array, and our data suggest a frequency-dependent dispersion measure.
\PSRf was discovered as a rotating radio transient, but is a nulling pulsar at 820\,MHz.
\PSRg is a new millisecond pulsar (MSP) in a 4-day orbit with a low-mass companion.  
Four of our pulsars already have published timing solutions, which we update in this work: the recycled wide binary \PSRc, the non-eclipsing black widow \PSRe, the disrupted recycled pulsar \psri, and the eclipsing binary MSP \psrl, which is in an 8.7\,hr orbit with a redback-mass companion.
\end{abstract}

\section{Introduction}\label{sec:intro}
\begin{figure*}
    \centering
    \includegraphics[width=1\textwidth]{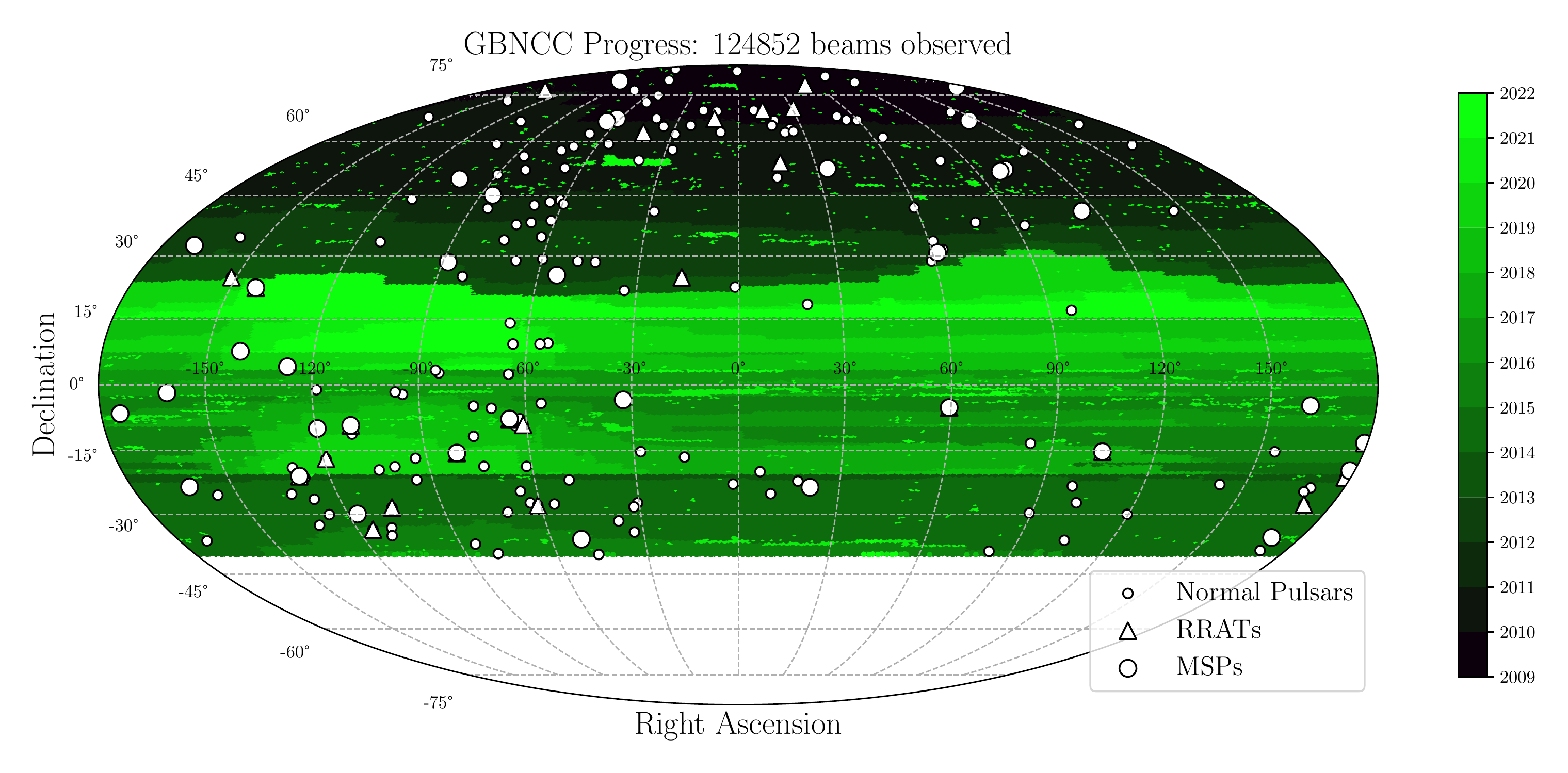}
    \caption{
    A skymap showing the GBNCC survey's sky coverage and its discoveries, which are differentiated between normal (``slow'') pulsars, rotating radio transients (RRATs), and millisecond pulsars (MSPs). 
    Each pointing is colored based on the date on which it was observed.
    }
    \label{fig:skymap}
\end{figure*}

The Green Bank North Celestial Cap (GBNCC) survey uses the Green Bank Telescope (GBT) to search for new pulsars at a radio frequency of 350\,MHz. 
Since the beginning of the survey in 2009, it has made 124,852 observations, each 120\,s in duration, covering the entire GBT sky (declination $\delta > -40\arcdeg$).
To date, GBNCC has discovered 194 pulsars.
The current sky coverage of the survey is shown in Figure~\ref{fig:skymap}.
In the coming months, the survey will be fully completed, as pointings which were rendered unusable by radio-frequency interference (RFI) are re-observed.

Survey observations use the Green Bank Ultimate Pulsar Processing Instrument \citep[GUPPI;][]{rdf+09} to sample the 
100\,MHz
bandwidth, which is split into 4096 frequency channels, once every 81.92\,\us. 
Data are processed on Compute Canada supercomputers at McGill University, using a searching pipeline which makes use of the pulsar searching software package \texttt{PRESTO}\footnote{\url{https://github.com/scottransom/presto}} \citep{sem+02}.
Candidate periodic and single-pulse signals are inspected by eye (often by undergraduate students), and promising candidates are followed up with the GBT.
Candidates which are confirmed as pulsars must be regularly observed over the course of a year to reach a phase-connected timing solution, which fully describes a pulsar's astrometric, rotational, and orbital parameters.

For a comprehensive overview and description of the GBNCC pulsar survey and its goals, as well as initial timing solutions for PSRs \psrc, \psre, \psri, and \psrl, which we are updating in this work, see \citet{slr+14}. 
Timing solutions for other GBNCC pulsar discoveries appear in \citet{ksr+12}, \citet{kkl+15}, \citet{kmk+18}, \citet{lsk+18}, \citet{acd+19}, \citet{amm+21}, and \citet{spp+23}.
The first GBNCC FRB discovery was reported in \citet{pck+20}. A census of the survey's discoveries and an analysis of its sensitivity can be found in \citet{mss+20}.
The survey maintains a website listing its discoveries and showing its current progress\footnote{\url{http://astro.phys.wvu.edu/GBNCC/}}, as well as a Github page\footnote{\url{https://github.com/GBNCC/data}} which provides published standard profiles, pulse times of arrival, and timing models from many of the above studies.

The main goal of the GBNCC survey is to discover new millisecond pulsars (MSPs).
Pulsars lose rotational energy over time, ``spinning down'' until radio emission ceases.
These old neutron stars sometimes go through a period of accretion from a binary companion.
This process, known as recycling, transfers angular momentum to the neutron star.
If recycling is allowed to proceed uninterrupted, pulsars are spun up to $\sim$ ms spin periods.
This process also reduces the strength of the pulsar's surface magnetic field \citep{acr+82}. 
Due to their extremely stable rotation, MSPs have high timing precision which can rival that of terrestrial atomic clocks \citep{hcm+12,hgc+20}.
This precision can be exploited to study a wide range of astrophysical phenomena, including tests of general relativity \citep[see, e.g.,][]{agh+18,ksm+21}, constraining the neutron star equation of state by measuring pulsar masses \citep{cfr+20,fcp+21}, and pulsar formation mechanisms and evolution.

By finding MSPs with sufficient timing precision, surveys like GBNCC are able to provide critical additions to pulsar timing array (PTA) experiments, which are Galaxy-sized gravitational wave (GW) detectors composed of many Earth--pulsar ``arms,'' sensitive to GWs at nanohertz frequencies.
The first hints of a nanohertz GW background may be present in the 12.5-year data set of the North American Nanohertz Observatory for Gravitational waves \citep[NANOGrav;][]{aab+21}, the North American PTA.
Similar hints of a signal are also present in the most recent International Pulsar Timing Array (IPTA) data release \citep{pdd+19,aab+22}, which combined data sets from PTAs in Australia (Parkes Pulsar Timing Array, PPTA), Europe (European Pulsar Timing Array, EPTA), and North America.
Eleven GBNCC discoveries have been observed by PTAs, including two pulsars in this analysis: \PSRe, which is currently observed by NANOGrav and the European PTA; and \PSRh, which was observed by NANOGrav until operations at the 305-m Arecibo radio telescope were suspended a few months prior to the tragic collapse of the telescope in December 2020.
When NANOGrav's timing program was transferred entirely to the GBT, observations of \PSRh were discontinued.

Additional goals for the GBNCC survey include discovering new nulling pulsars and rotating radio transients \citep[RRATs;][]{mll+06}; exotic binary systems such as double neutron star (DNS) systems, black widows and redbacks \citet{fst+88,rms+11}; and studying the Galactic pulsar population as a whole. 
Black widows and redbacks, collectively known as ``spider'' binaries, are MSPs in short, $P_\mathrm{b} \lesssim 1$\,day orbits. 
In these systems, the companion is ablated by the energetic pulsar wind, releasing ionized material into the system, which smears and delays the pulsar's radio pulses or causes radio eclipses \citep{pbc+18}.
Spiders have a bimodal distribution of companion masses, with black widows having $M_\mathrm{c} \lesssim 0.05$\,\Msun and redbacks having $M_\mathrm{c} \gtrsim 0.1$\,\Msun \citep{cct+13}.

Recycled pulsars which are neither MSPs nor in binary systems are known as disrupted recycled pulsars (DRPs), defined by \citet{blr+10} as isolated pulsars in the Galactic disc (i.e., not in a globular cluster, where many-body interactions can easily disrupt binary systems) with spin periods of $P > 20$\,ms, and low surface magnetic fields, $B_\mathrm{surf} < 3 \times 10^{10}\,\gauss$.
These properties suggest that such a DRP was in the process of accreting from a binary companion when that companion underwent a supernova explosion, imparting a kick that disrupted the binary.
This may result in larger space velocities for DRPs compared to other pulsar populations, such as DNS binaries \citep{lma+04}.
This picture of pulsar evolution can be tested by, e.g., comparing the relative numbers and/or space velocities of DNS systems and DRPs \citep{kmk+18}.

We note that one of the pulsars in this analysis, \PSRm, was reported as a discovery in the HTRU-North survey\footnote{Several pulsars were listed as ``co-discoveries'' with the GBNCC survey; this was not the case with \PSRm, though it was published as a GBNCC discovery in \citet{slr+14}.}, a pulsar survey at 1.36\,GHz with the Effelsberg radio telescope \citep{bck+13}.
That paper also includes a timing solution for this pulsar with parameters consistent with, and comparable in precision to, those presented in this work. 
In this work, we present our own independent pulse profiles, flux density measurements, and timing solution for this pulsar.

In Section~\ref{sec:observations}, we describe our timing observations of 21 pulsars. 
We present pulse profiles, estimated flux densities, and spectral indices in Section~\ref{sec:flux} and describe our timing analysis in Section~\ref{sec:timinganalysis}. 
In Section~\ref{sec:results}, we present our timing solutions and discuss some individual systems.
We conclude with Section~\ref{sec:conc}.

\section{Pulsar Timing Observations}\label{sec:observations}
The discoveries and initial timing follow-up observations of PSRs \psrc, \psre, \psri, and \psrl were detailed in \citet{slr+14}.
We refer to that work for detailed descriptions of those observations (which were made with the GBT at 350, 820, 1500, and 2000\,MHz), describing all new timing observations here.
The pulsars in this analysis were each discovered as periodicity candidates in GBNCC survey observations, except for \PSRf, which was discovered in a search for single pulses in the 350\,MHz GBT Drift-scan survey \citep{kkl+15}.

After confirmation with the GBT at 350\,MHz, many of the pulsars in this analysis were used as test sources during regular survey observing.
Observations of known pulsars as test sources are performed during each survey observing session to ensure data quality and monitor the RFI environment.
These test scans use the same observing set up as usual survey observations: the 100\,MHz
bandwidth, centered at 350\,MHz, is split into 4096 frequency channels, with a sampling time of 81.92\,\us.
A small number of test source observations used in our timing analysis were made with the newer VEGAS backend instead of GUPPI, using an identical setup.

At 350\,MHz, the GBT beam has a FWHM of 36\arcmin, so the sky positions of recently-confirmed pulsars are not precisely known.
This can cause difficulty in reaching a timing solution, and can significantly reduce the signal-to-noise ratio (S/N) of timing observations, as the 820\,MHz beam is only $7\arcmin$ wide.
To ameliorate this issue, pulsar positions were refined using the traditional gridding technique.
For each pulsar, six observations were made sequentially, each at 820\,MHz, with 200\,MHz
bandwidth split into 2048 channels and 40.96\,\us 
time resolution.
These gridding beams were distributed evenly across the 350\,MHz discovery beam.
The varying S/N at each sky location allows the inference of an improved pulsar position.
As long as the gaps between our timing follow-up observations of a pulsar and any discovery, test source, and gridding observations were short enough, such that phase connection could be maintained, we used them in our timing analysis.

After localization, pulsars were observed once monthly for a year at the GBT (project code 15A-376; PI: L.~Levin) at 820\,MHz. 
Each pulsar had at least one period of high cadence observing, with 4 -- 5 observations made within a period of one week, to facilitate phase connection and the solving of orbital parameters, if applicable.
Timing solutions were not available for many pulsars at the outset of the timing campaign, so the majority of our observations were not coherently folded or dedispersed (``search-mode'' data, which have the same configuration as the gridding scans described previously). 
For a few pulsars, a suitable timing ephemeris was available, so data were coherently dedispersed to the correct DM and folded on the pulsar's spin period (``fold-mode'' data), using 2048 phase bins and 10\,s
sub-integrations.
Raw fold-mode data contained polarization information, and polarization calibration scans were taken, but we leave a polarization analysis of these pulsars to a future work.

Some pulsars were also observed under two similar GBT timing campaigns, each lasting $\sim$1\,yr: one using the same setup at 820\,MHz (project code 17B-285; PI: J.~Swiggum), the other observing at 350\,MHz (project code 16A-343; PI: M.~DeCesar).
Once again, search mode was used for some pulsars, with a setup at 350\,MHz identical to GBNCC survey observations, and fold mode for others.
With the 350\,MHz receiver, fold mode uses 128 frequency channels and a 1.28\,\us
sampling time.

We observed \PSRc with the Low Frequency Array (LOFAR) at 149\,MHz, under project number LC0\_002.
These observations are described in detail in \citet{lsk+18,kvh+15}. 
Data were recorded using 78.125\,MHz of bandwidth split into 400 subbands, each split into 16 channels with a sampling time of 327.68\,\us.

We also used the Long Wavelength Array (LWA) to observe \PSRh approximately once every three weeks between MJDs 56863 (2014 September) and 57869 (2017 April).
Observations were made with the LWA Station 1 (LWA1).
The LWA1 \citep{etc+13} is capable of forming four independently-steerable beams, each with two independently-selectable center frequencies with up to 19.6\,MHz of bandwidth each (due to rolloffs in sensitivity towards the edges of the band, the usable bandwidth per tunable center frequency is $\sim$ 16\,MHz).
Most of our observations used two beams, one with center frequencies at 35.1 and 49.8\,MHz, and another with 64.5 and 79.2\,MHz, using the maximum bandwidth available.
On some epochs, \PSRh was only observed with one beam, and thus at only two frequencies, but these were chosen such that each portion of the band was observed a total of 41 times, except for 49.8\,MHz, which was observed 40 times.
Data were coherently dedispersed and folded (30\,s
duration sub-integrations) using a real-time spectrometer with a sampling time of 81.08\,\us.
Each beam had 1024 frequency channels available, so the bands corresponding to each center frequency were each split into 512 channels, resulting in a channel bandwidth of $\approx$ 38.3\,kHz. 

Observations of \PSRh by the NANOGrav PTA, using the 305-m William E.~Gordon radio telescope at Arecibo Observatory (AO), were made available to GBNCC.
This is in accordance with the data-sharing agreement between major pulsar surveys and PTAs, whereby the surveys share timing ephemerides of high-timing-precision MSP discoveries with PTAs, and the PTAs share timing products with the surveys.
AO observations of this pulsar were taken in the same manner as described in \citet{abc+15}, but we summarize them here.
Observations were made at $\sim$ monthly cadence, at center frequencies 430 and 1380\,MHz, with an observation with one receiver followed immediately by an observation with the other within $\sim$1\,hr, accompanied by measurements of pulsed noise diode signals to calibrate the polarization response of the receiver.
Most observations were 19\,min in duration, with a few as short as 10\,min and the longest at 40\,min.
Data were recorded by the Puerto Rican Ultimate Pulsar Processing Instrument (PUPPI; nearly identical to GUPPI) pulsar backend with a sampling time of 64\,\us, and 1.5625\,MHz-wide frequency channels, with a bandwidth of 24 and 800\,MHz for the 430 and 1380\,MHz 
receivers, respectively.
Observations were folded and dedispersed coherently using the pulsar ephemeris and DM, resulting in data products with 
10\,s 
sub-integrations and 2048 pulse phase bins.

\section{Pulse Profiles, Flux Densities, and Spectral Indices}\label{sec:flux}
\begin{figure*}
\centering
\includegraphics[width=1\textwidth]{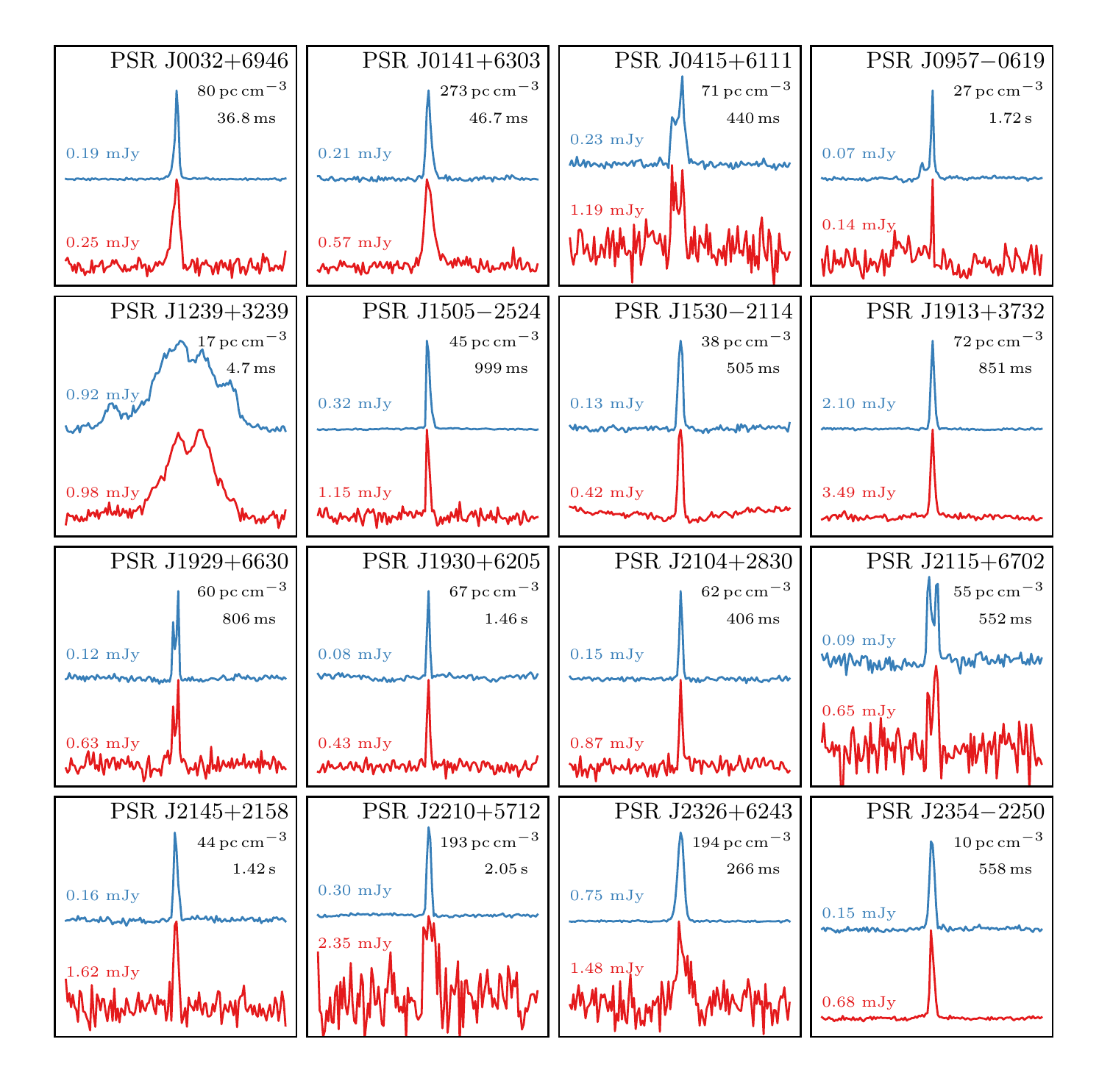}
\centering
\caption{Summed pulse profiles for 16 of the pulsars in this analysis. 
Each profile has been normalized to the same peak height, shows the emission from one full rotation of the pulsar, and is divided into 128 bins. 
They were generated using all GBT data which were within the degradation tolerances described in Section~\ref{sec:flux}. 
We show 350\,MHz
profiles in red and 820\,MHz 
profiles above them in blue.
Below each pulsar's name, we give its DM and spin period.
}
\label{fig:profs1}
\end{figure*}

\begin{figure*}
\centering
\includegraphics[width=1\textwidth]{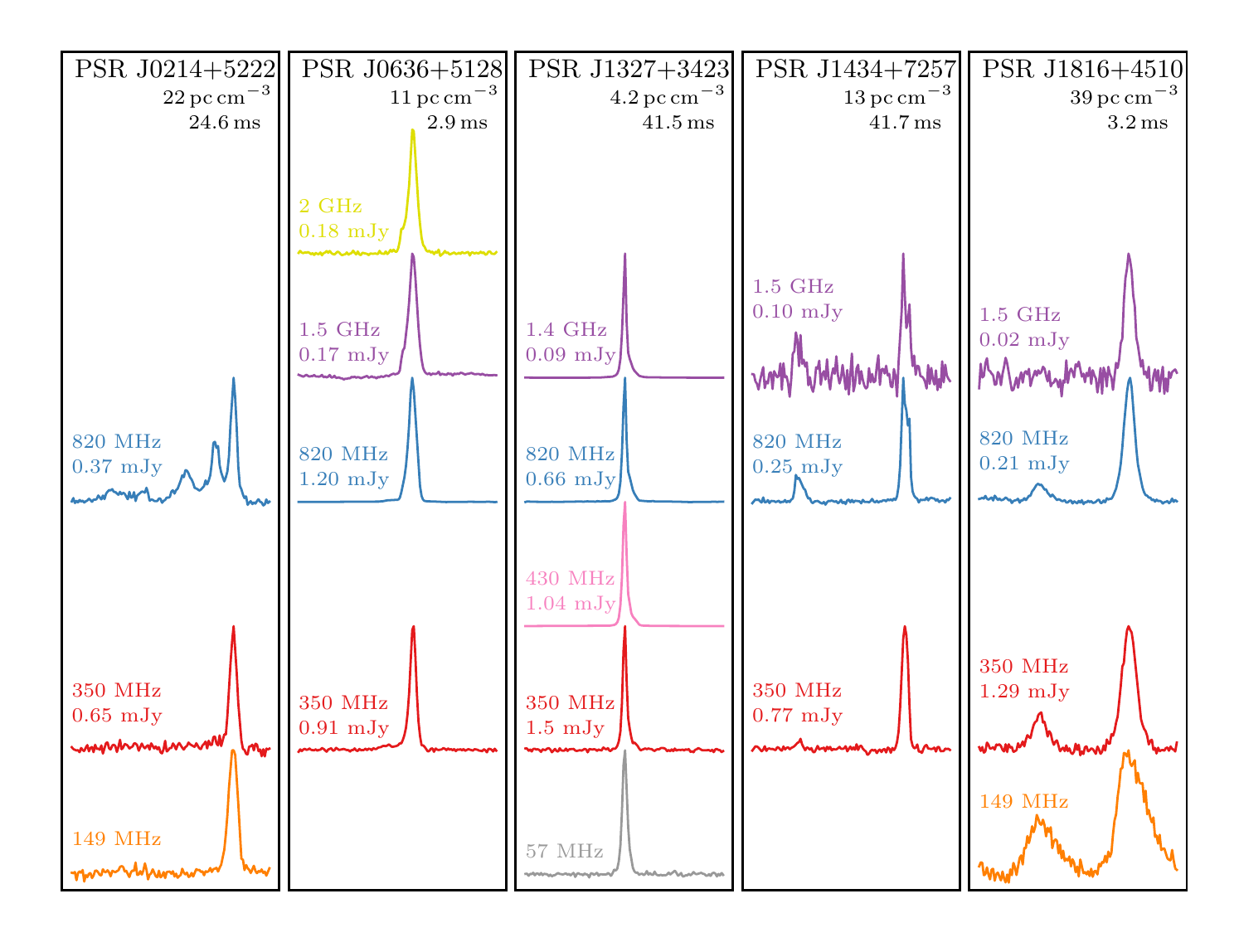}
\centering
\caption{Summed pulse profiles for the five pulsars in this analysis which were observed at additional telescopes/frequencies. 
Each profile has been normalized to the same peak height, shows the emission from one full rotation of the pulsar, and is divided into 128 bins. 
They were generated from all data which were within the degradation tolerances described in Section~\ref{sec:flux}.
The telescopes used were LWA1 (57\,MHz), LOFAR (149\,MHz), GBT (350, 820, 1500, and 2000\,MHz), and AO (430 and 1380\,MHz).
We also show the 149\,MHz
LOFAR profile of \PSRl, which is the same profile shown in \citet{slr+14}.
The center frequency of the observations used to generate each profile and the estimated flux density $S_\nu$ at that frequency are listed to the left of each profile.
We did not estimate $S_\nu$ from LOFAR profiles, and we estimated $S_\nu$ separately for each LWA1 subband, so we do not list $S_\nu$ estimates beside those profiles.
Below each pulsar's name, we give its DM and spin period.}
\label{fig:profs2}
\end{figure*}

We estimate flux densities $S_\nu$ for each pulsar at each observing band's central radio frequency $\nu$ using the radiometer equation as presented in \citet{lk+04}:
\begin{equation}\label{eq:radiometer}
S_\nu = \beta\,\frac{(\mathrm{S/N})\,T_\mathrm{sys}}{G\,\sqrt{n_\mathrm{p}\,t_\mathrm{int}\,\Delta \nu}}\,\sqrt{\frac{\delta}{1-\delta}}.
\end{equation}
Here, $\beta=1.3$ is a degradation factor due to digitization, S/N is the signal-to-noise ratio, $T_\mathrm{sys}$ is the system temperature, $G$ is the telescope gain, $n_\mathrm{p}=2$ is the number of polarizations, $t_\mathrm{int}$ is the total integration time on source, $\Delta\nu$ is the effective bandwidth, and $\delta$ is the pulse duty cycle.
S/N was measured from the summed pulse profiles shown in Figures \ref{fig:profs1} and \ref{fig:profs2}.
We calculated equivalent widths $W_\mathrm{eq}$ for each pulse profile, defined in \citet{lk+04} as the width of a boxcar pulse with the same area and peak height as the pulse profile. The duty cycle is then $\delta = W_\mathrm{eq}/P$. 
We report $\delta$ and $W_\mathrm{eq}/P$ for each pulsar in Table~\ref{tab:duty}; we did not add together the LWA1 bands and instead report distinct measurements at 35.1, 49.8, 64.5, and 79.2\,MHz for \PSRh in Table~\ref{tab:LWA}.


\input{duties-widths}\label{tab:duty}

\input{LWA.tex}\label{tab:LWA}

\vspace{-13.5mm}
To ensure we account for persistent sources of RFI and rolloffs in sensitivity at the edges of the band, we assumed $\Delta\nu = 90$\% of the true observing bandwidth.
We reduced this to 75\% for AO observations at 1380\,MHz due to increased RFI.
In 2014, a new source of strong, persistent RFI rendered GBT data in the range 360 -- 380\,MHz unusable.
This has caused a $\approx$20\% reduction in the effective bandwidth of the 
350\,MHz 
receiver.
For pulsars with 
350\,MHz 
observations both before and after the change, we took $\Delta\nu_{350} = 80\,\MHz$.
Observations of \PSRb at 
350\,MHz
only occurred after this source of RFI appeared, so we took $\Delta\nu_{350} = 70\,\MHz$.

The system temperature $T_\mathrm{sys} = T_\mathrm{rec} + T_\mathrm{sky}$, where $T_\mathrm{rec}$ is the receiver temperature and $T_\mathrm{sky}$ is the position-dependent sky temperature, including contribution from the cosmic microwave background.
Referencing the GBT Proposer's Guide\footnote{\url{https://www.gb.nrao.edu/scienceDocs/GBTpg.pdf}, see Table 3 for $G$ and Figure 3 for $T_\mathrm{rec}$}, we took $T_\mathrm{rec}$ to be 23, 22, 20, and 18\,K, at 350, 820, 1500, and 2000\,MHz, respectively.
From the Arecibo 305-m telescope User's Guide\footnote{\url{http://www.naic.edu/~astro/User_Guide_2020.pdf}, see Table 3}, $T_\mathrm{rec} = 35$ and 25\,K at 430 and 1380\,MHz, respectively.
We used \texttt{pyGDSM}\footnote{\url{https://github.com/telegraphic/pygdsm}}, a Python interface for \citet{ztd+17}'s global sky model of diffuse radio emission, to obtain $T_\mathrm{sky}$ at each relevant frequency at the sky location of each pulsar.

The telescope gain, $G$, is in practice a function of the angle between telescope boresight and the true position of the pulsar, $\theta$.
We modeled $G(\theta)$ as a Gaussian function with a maximum value equal to the boresight gain $G(0)$, and FWHM equal to that of the telescope beam.
For AO and GBT, we used values for $G(0)$ found in the aforementioned user guides: 2~K\,Jy$^{-1}$ for the GBT, and 11/10.5~K\,Jy$^{-1}$ for AO at 430/1380\,MHz, respectively.
For LWA1 observations, we instead substituted the System Equivalent Flux Density, $\mathrm{SEFD} = T_\mathrm{sys}/G(0) \sim 20$\,kJy \citep[Fig. 12 in][using 6$\arcdeg$ as the zenith angle, which was typical for our observations of \PSRh]{etc+13}, into Equation~\ref{eq:radiometer}.

Following \citet{spp+23}, we used this simple model of the telescope beam to estimate degradation factors for each observation, $\mathrm{DF} = G(\theta)/G(0)$.
Approximately half of our pulsars had a high fraction of observations which were over the separation threshold used in \citet{spp+23}, that being the angular separation from a pulsar's timing position which would cause $>$10\% degradation in S/N.
Therefore, for each pulsar/band combination, we added to this the median of the observations' separations to reach our final separation threshold.
For example, at 820\,MHz, GBT observations separated from the pulsar's true position by 3\arcmin\ will have DF = 0.9 (10\% degradation).
The majority of our 21 observations of \PSRk at 820\,MHz are at $>$3\arcmin\ separations from the pulsar's timing position, with a median separation of 5.4\arcmin; our threshold is then 8.4\arcmin.
This retained the majority of the observations of this pulsar, while disregarding six which have significantly-lower DF.
Any observations with separations higher than the thresholds were not used to create profiles or estimate flux densities, though we did not necessarily discard them from our timing analysis.

\begin{figure}
\centering
\includegraphics[width=0.5\textwidth]{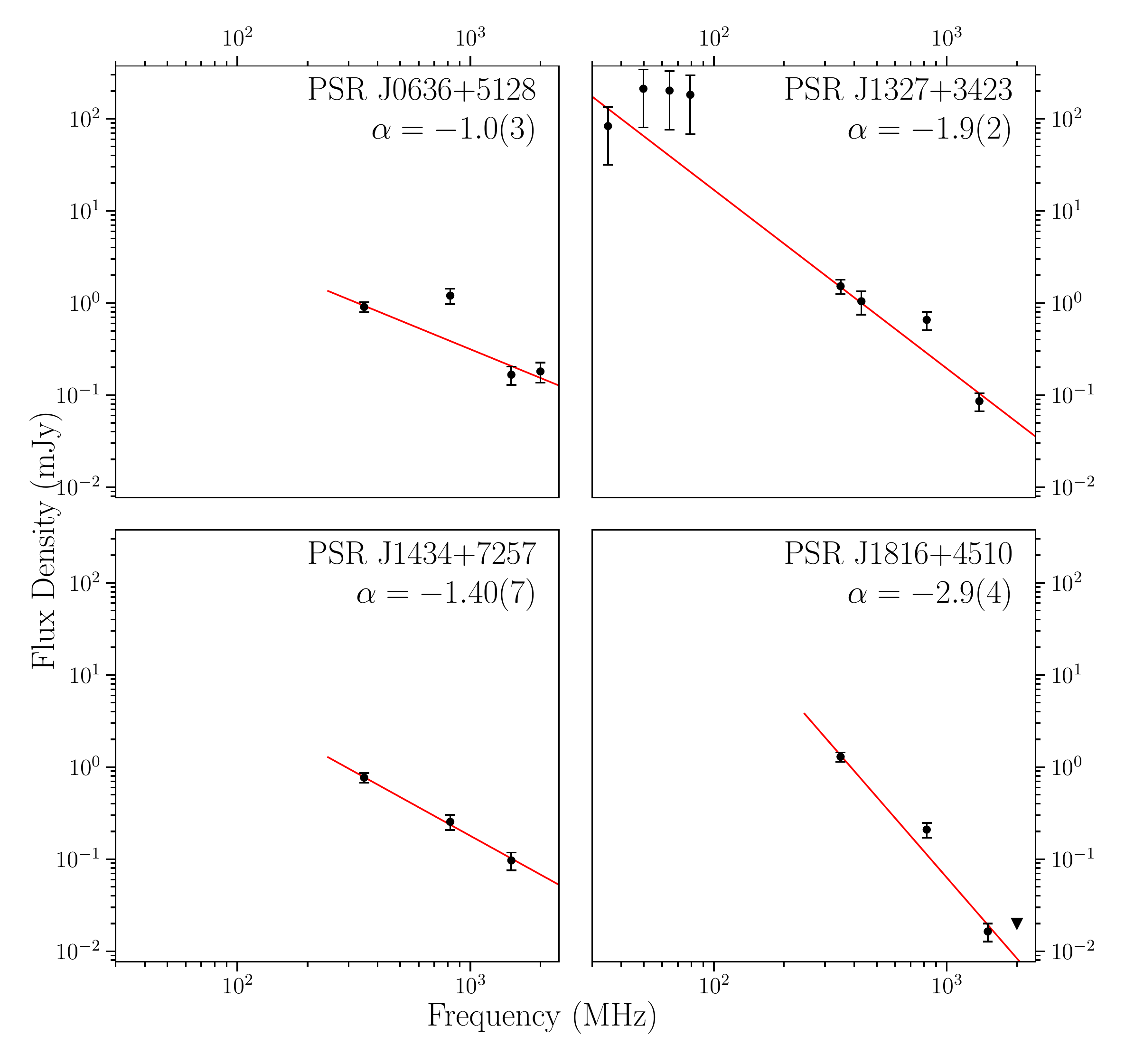}
\centering
\caption{Flux density ($S_\nu$) measurements and best-fit power law spectra for PSRs \psre, \psrh, \psri, and \psrl. The triangle in the bottom-right plot represents an upper limit based on our lack of a detection of \PSRl at 2\,GHz.}
\label{fig:spectra}
\end{figure}

We assume each pulsar's spectrum follows a power law with spectral index 
$\alpha$, $S_\nu \propto \nu^\alpha$.
For pulsars with estimates at more than two bands, we performed a least-squares fit to the flux density measurements in log-log space. 
Best-fit power laws are shown in Figure~\ref{fig:spectra}.
We report total integration times used to generate profiles, flux density estimates, and spectral indices in Table~\ref{tab:flux}.
Flux densities measured for \PSRh from LWA1 observations are presented separately in Table~\ref{tab:LWA}.

\input{flux.tex}\label{tab:flux}

\vspace{-8.5mm}
We estimated uncertainties using standard error propagation, assuming uncertainties in $\Delta\nu$, $T_\mathrm{sys}$, and $\delta$ as follows. 
Day-to-day changes in $T_\mathrm{sys}$ on the level of a few K are expected, so we assumed $\sigma_{T_\mathrm{sys}} = 5\,\mathrm{K}$.
We assumed an uncertainty in $\delta$ equal to one phase bin, or $\sigma_\delta = 1/128$, as they were chosen manually.
Transient sources of RFI can alter the effective bandwidth of individual observations, so we assumed $\sigma_{\Delta\nu} = 0.1\Delta\nu$. 
For pulsars with 
350\,MHz 
observations both before and after the aforementioned drastic change to the RFI environment which occurred in 2014, we increased $\sigma_{\Delta\nu}$ to 20\,MHz to reflect the change.

As discussed earlier, \PSRf was detected in the GBT 350\,MHz Drift-scan survey.
It was not detected in the GBNCC survey observation closest to its timing position, which was severely affected by RFI.
In order to estimate $S_\nu$ at 350\,MHz for this pulsar, and thus $\alpha$, we folded the discovery drift-scan observation on the pulsar ephemeris we obtained through our timing analysis.
This yielded a 
350\,MHz
profile which was weak, but sufficient to estimate $S_\nu$.
The drift-scan observation was taken in the same setup as the GBNCC survey observations described in Section~\ref{sec:observations}, but with only 50\,MHz of bandwidth. 
The declination of the drifting telescope beam was $-06\arcdeg\, 17\arcmin\, 20\farcs52$, sufficiently close to the true declination of the pulsar for the changing separation between the two not to significantly impact sensitivity during the 2.6-minute scan.

\PSRl was not detected in three S-band observations with the GBT, $\sim$17 minutes total integration time.
We place an upper limit at 2\,GHz of $S_\nu < 0.02$\,mJy for this pulsar, assuming $\delta = 0.06$ and $\mathrm{S/N} < 6$.

\begin{figure}[t!]
\centering
\includegraphics[width=0.49\textwidth]{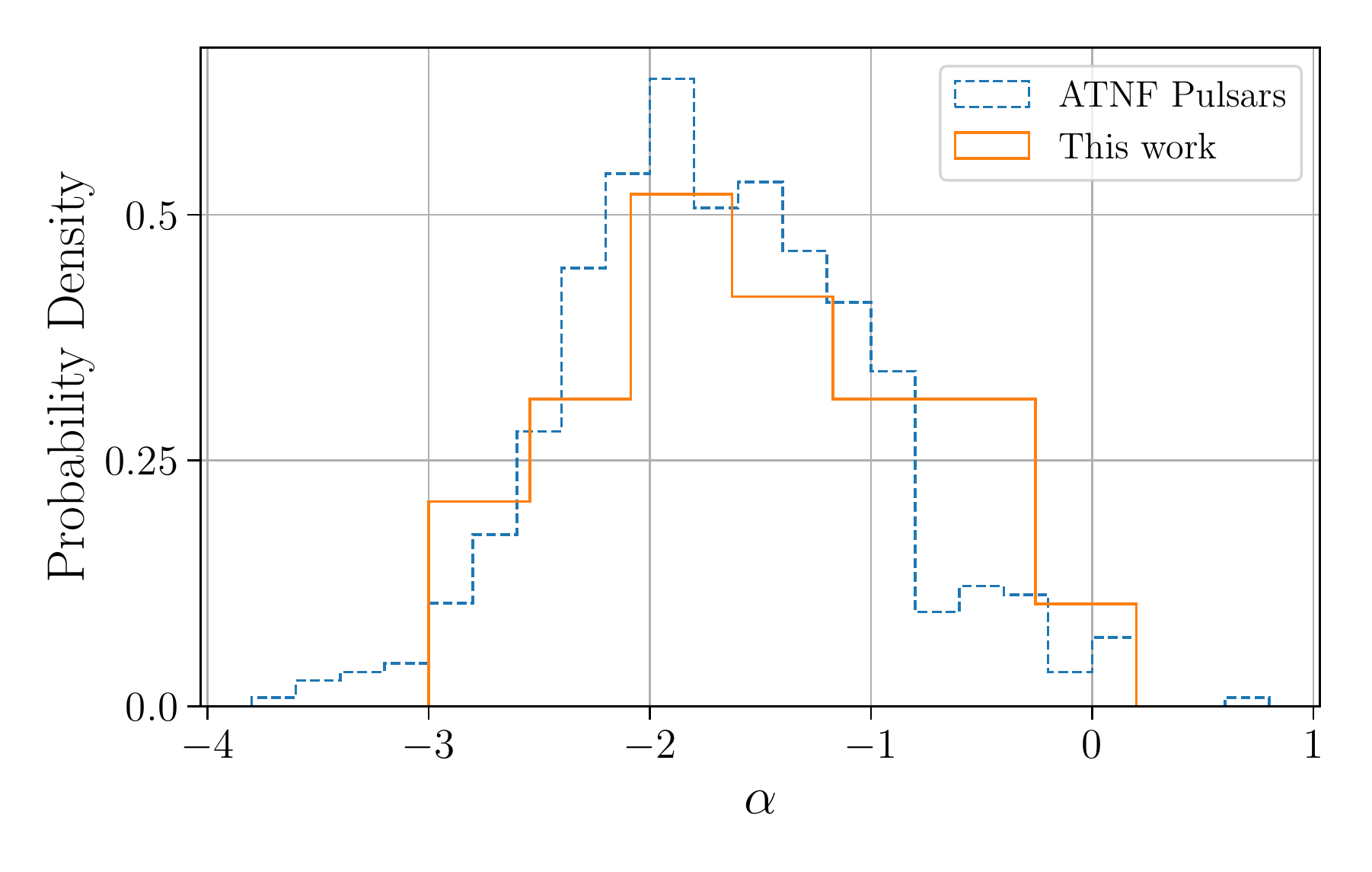}
\centering
\caption{
Histograms showing the distribution of pulsar spectral indices ($\alpha$) for pulsars in the ATNF catalog (dashed line) and those in this work (solid line). 
We calculated spectral indices for pulsars in the ATNF catalog using flux-density measurements at 400 and 1400\,MHz.
Visually, there appears to be an excess of flatter-spectrum pulsars in this work with $-1 \lesssim \alpha \lesssim 0$, but a Kolmogorov--Smirnov test does not indicate that the two samples are drawn from different underlying distributions.}
\label{fig:alphahist}
\end{figure}

We note that several pulsars in our sample have relatively flat spectra, with $\alpha$ even consistent with zero in a few cases. 
We selected each pulsar in the ATNF pulsar catalog\footnote{\url{https://www.atnf.csiro.au/people/pulsar/psrcat/}} \citep{mht+05} with listed flux densities at both 400 and 1400\,MHz (\texttt{S400} and \texttt{S1400}) and calculated spectral indices using those measurements.
Histograms comparing the distribution of spectral indices in ATNF versus those of the pulsars in this work are shown in Figure~\ref{fig:alphahist}.
We compared the two distributions using a statistical Kolmogorov--Smirnov test, and found that we could not refute the null hypothesis ($p \approx 0.5$) that the two samples are drawn from the same underlying distribution.
This is perhaps contrary to the natural expectation that a low-frequency survey would discover a greater number of steeper-spectrum sources.

\section{Pulsar Timing Analysis}\label{sec:timinganalysis}
For information about the discoveries and initial timing analysis of PSRs \psrc, \psre, \psri, and \psrl, see \citet{slr+14}.
When timing data first became available, each scan was processed using \texttt{PRESTO}, both to obtain an initial set of times of arrival (TOAs) and to note any changes in apparent spin periods due to possible Doppler shifting from binary motion.
Only 350 and 
820\,MHz
observations with the GBT were used for this purpose.
We used \texttt{rfifind} to mask RFI and \texttt{prepdata} to produce timeseries at the pulsar's known DM, which were searched with \texttt{accelsearch} for periodicity candidates with periods close to the discovery value.
Finally, the raw data were folded with \texttt{prepfold}, which also searches for an improved spin period and period derivative.

Time-varying spin periods were noticed for PSRs \psra and \psrg, and preliminary sets of Keplerian parameters were obtained by performing a least-squares sinusoidal fit to the spin periods.
These became the starting orbital parameters for these pulsars' timing models.

We created standard profiles from the folded data using \texttt{PRESTO}'s \texttt{pygaussfit.py} to fit Gaussian components to the highest-S/N profile for each pulsar.
If useful data were available at both 350 and 820\,MHz, a separate standard profile was created at each frequency, and the two were aligned.
We then cross-correlated these standard profiles in the Fourier domain with the folded data \citep{tjh+92} using \texttt{get\_TOAs.py} to obtain TOAs.
We created three TOAs per 5 -- 15 minute observation to allow fitting for spin frequency at each epoch, enabling phase connection across day--week-long timespans at first, and eventually across each pulsar's entire data set.
These initial phase-connected timing solutions were obtained separately by several of the authors, who used either the \textsc{tempo}\footnote{\url{http://tempo.sourceforge.net}} or \tempotwo\footnote{\url{https://www.atnf.csiro.au/research/pulsar/tempo2/}} timing software.

Henceforth, we discuss the data processing in the context of a single pulsar's set of timing observations.
Raw data were folded on the new pulsar ephemeris, using \texttt{{fold\_psrfits}}\footnote{from\;\texttt{psrfits\_utils}\;(\url{https://github.com/demorest/psrfits_utils})} to fold search-mode data on the pulsar's spin period, resulting in 
10\,s
sub-integrations.
\psrchive\footnote{\url{http://psrchive.sourceforge.net/}}, a suite of pulsar data analysis software, was used for all further data processing.
Any data containing polarization information were first reduced to total intensities.
RFI was excised automatically using \texttt{paz} both before and after averaging, or ``scrunching'', to 128 frequency channels, in order to zap RFI from both single frequency channels and larger portions of the band as thoroughly as possible without removing useful data. 
Then, each frequency-scrunched observation was examined by eye and any remaining RFI was removed with \texttt{pazi}. 
Where applicable, any periods of nulling at the beginning or end of an observation were removed by extracting the appropriate sub-integrations using \texttt{pam}.

We then used \texttt{psradd} to phase-align observations and sum them to create an average profile for each band. 
To each of these, we used \texttt{paas} to fit Gaussian components, resulting in noise-free template profiles. 
Each observation was then scrunched in time and frequency to achieve the desired number of sub-integrations and subbands.
These numbers were generally 2 -- 5, with the exact number of each being chosen to both enable a determination of DM and avoid degrading S/N below 6. 
We set the maximum sub-integration length for \PSRe at 2.5\% of its 
1.6\,hr
orbital period, to minimize any smearing within sub-integrations due to Doppler shifts caused by orbital motion. 
Two full-orbit observations of \PSRl were split into many 9 -- 10 minute sub-integrations. 
Once a timing solution was initially reached, lower-S/N observations were sometimes fully scrunched in time so that more subbands could be used, in order to better constrain DM.

Data from AO observations of \PSRh were reduced according to the usual NANOGrav procedure described in \citet{abc+15}, from RFI removal and flux and polarization calibration, to time- and frequency-scrunching to sub-integrations up to 30 minutes long, with 64 subbands.
From that point, we followed the steps laid out above for the creation of template profiles.
For consistency with the AO data, we fully time-scrunched our 820\,MHz GBT observations of this pulsar and divided them into 32 subbands.

After scrunching, and ensuring standard profiles were correctly aligned (see below), one TOA per subband for each sub-integration was calculated using \texttt{pat}.
TOAs were later excised from the timing analysis if they were outliers, due to either very large uncertainty or corresponding timing residual $\delta t = \mathrm{TOA_{meas}} - \mathrm{TOA_{pred}}$, where $\mathrm{TOA_{meas}}$ is the measured TOA and $\mathrm{TOA_{pred}}$ is the corresponding TOA predicted by the pulsar's timing model. 
316 out of 8726 total TOAs were flagged as outliers in this manner. 

We eliminated undesired offsets between TOAs from different telescopes/receivers using three methods. 
First, the template profiles were aligned to the same reference phase using \texttt{pas}.
Second, for TOAs obtained using the GUPPI or VEGAS backends, known timing offsets were removed by adding \texttt{TIME} flags to TOA (\texttt{tim}) files. 
Timing offsets between different observing modes and receivers at the GBT were determined using \texttt{guppi\_offsets}\footnote{\url{https://github.com/demorest/guppi_daq}} for GUPPI observations: e.g. search-mode 
350\,MHz 
data is offset from fold-mode 
820\,MHz 
data by 78.08\,\us; and \texttt{vpmTimingOffsets.py} (a command on GBO computers) for VEGAS.
Third, for pulsars with observations at different telescopes, timing offsets between different observatories and receivers were fit for using JUMPs (an arbitrary phase offset, which is fit in the timing model) in the pulsar parameter (\texttt{par}) files.
Due to missing data files, we were unable to reproduce TOAs corresponding to some older observations of PSRs \psrc and \psrl, and instead used the same TOAs which were used in \citet{slr+14}.
In order to account for the different folding ephemerides and standard profiles used to generate these old TOAs, we also fit JUMPs for these older TOAs, doing so separately for, e.g., 350 and 
820\,MHz.

Timing parameters were then fit for iteratively using \tempotwo, using the DE440 solar system ephemeris and TT(BIPM2021) time standard.
The introduction of new timing parameters, such as proper motion and parallax, was tested during this process.
If it was not obvious whether a new parameter was significant, we used a statistical $F$-test to compare the $\chi^2$ of the fit with and without the new parameter, only including it in the timing model if it passed 3-$\sigma$ confidence with $\alpha < 0.0027$.
After a timing solution was reached, the data were refolded using the updated ephemeris, TOAs were re-created using the same method as before, and timing parameters were fit once again using this final set of TOAs.
This last step was necessary because the pulsar ephemeris originally used to fold the data was necessarily incorrect, and this could introduce errors into the final timing solution.

\subsection{Timing of Binary Pulsars}

Pulsar binary orbits are characterized by, at minimum, five Keplerian parameters: the orbital period $P_\mathrm{b}$, projected semimajor axis $x = a\sin i/c$ (where $a$ is the semimajor axis and $i$ is the inclination of the orbit, $i=90\arcdeg$ being edge-on when viewed from Earth), eccentricity $e$, longitude of periastron $\omega$, and time of periastron $T_0$.
For low-eccentricity orbits, there are high covariances between $\omega$ and $T_0$, resulting in high uncertainties \citep{ELL1}.
All of the binary pulsars in this work have eccentricities which are low enough to cause these high covariances.

Therefore, we used the \citet[ELL1]{ELL1} binary model, which uses an approximation for the Roemer delay (\dre) and parameterizes the orbit in terms of the time of ascending node,
\begin{equation}\label{eq:tasc}
    T_\mathrm{asc} = T_0 - P_\mathrm{b}(\omega/2\pi), 
\end{equation}
and the Laplace-Lagrange parameters,
\begin{equation}\label{eq:e1e2}
    e_1 = e\sin\omega\ \ \mathrm{and}\ \ e_2 = e\cos\omega.
\end{equation}
\tempotwo's implementation of the ELL1 model contains terms for \dre~up to first order in $e$. 
This is sufficient for pulsars with $\delta t_\mathrm{RMS}/N_\mathrm{TOA}^{1/2} \gtrsim xe^2$, but PSRs \psra and \psrc do not satisfy this requirement.
We attempted to use the \citet{bt+76} model for these pulsars, which uses $e$, $T_0$, $\omega$, and the full expression for \dre.
The timing solutions seemed serviceable, but covariances between $T_0$ and $\omega$ were quite high.

Fortunately, the second-order terms for \dre, calculated by \citet{zdw+19}, have been implemented in (current version, not the latest release of) the pulsar timing software PINT\footnote{\url{https://github.com/nanograv/PINT}} \citep{lrd+21}.
These are sufficient for \PSRa, but \PSRc has $xe^3 > \delta t_\mathrm{RMS}/N_\mathrm{TOA}^{1/2}$.
Therefore, to reach an accurate solution for this pulsar's binary motion, it is necessary to extend the approximation for the Roemer delay to third order in $e$:
\begin{widetext}
\begin{align}
    \Delta_\mathrm{R} \simeq\ &x\bigg(\sin\phi + \frac{e_2}{2}\sin 2\phi - \frac{e_1}{2}\cos 2\phi\bigg)
    - \frac{x}{8}\Big(5e_2^2\sin\phi - 3e_2^2\sin 3\phi -2e_1e_2\cos\phi
    + 6e_1e_2\cos 3\phi 
    + 3e_1^2\sin\phi + 3e_1^2\sin 3\phi\Big) \nonumber \\
    &- \frac{x}{12}\Big(
    5e_2^3\sin 2\phi 
    + 3e_1^2e_2\sin 2\phi 
    - 6e_1e_2^2\cos 2\phi 
    - 4e_1^3\cos 2\phi 
    - 4e_2^3\sin 4\phi 
    + 12e_1^2e_2\sin 4\phi 
    + 12e_1e_2^2\cos 4\phi 
    - 4e_1^3\cos 4\phi
    \Big) \label{eq:roemer}
\end{align}
\end{widetext}
(note that we have corrected here an index swap in the first-order terms present in Equation 1 of \citet{zdw+19}), where $\phi$ is the orbital phase used in the ELL1 model, written in radians as 
\begin{equation}\label{eq:phase}
    \phi(t) = (t - T_\mathrm{asc})(2\pi/P_\mathrm{b}).
\end{equation}
This expression is sufficient for \PSRc, since $xe^4 \simeq 0.14$\,\us.
We implemented this expression in PINT, which we used to produce final timing solutions for these two pulsars.
Note that we did \textit{not} use this ``third-order ELL1'' model for PSRs \psre, \psrg, or \psrl; the first-order ELL1 model in \tempotwo is sufficient for these systems.

\subsection{Accounting for DM Variations}\label{sec:dmx}
Variations in DM are expected to occur on $\sim$monthly timescales due to the line of sight to the pulsar traversing regions with different electron densities. These effects are on the order of $10^{-(3-4)}$\;\dmu \citep{jml+17}, significantly lower than the precision of some of our DM measurements. 
One method of modeling variations in DM in the timing model is fitting a piecewise constant function called DMX. 
A fixed reference value for DM is chosen, and each DMX parameter describes an offset from that value at a particular epoch (a grouping of observations, usually 1 -- 7 days in length).

For most of our pulsars, applying the DMX model is unnecessary, so we simply note that any reported DM uncertainties $\lesssim 0.001$\,\dmu are likely underestimated. 
However, DM variations could lead to a meaningful signature in the residuals for our most precisely-timed pulsars.
The dispersion delay between two frequencies $\nu_1$ and $\nu_2$ (in MHz) is given by
\begin{equation}
    \Delta t \simeq 4.15 \times 10^6\,\mathrm{ms} \times (\nu_1^{-2} - \nu_2^{-2}) \times \mathrm{DM}
\end{equation}
\citep{lk+04}. 
For a change $\sim 10^{-3}$\;\dmu, the corresponding delay compared to infinite frequency at 820 MHz is $\sim$ 6.2\,\us, comparable to $\delta t_\mathrm{RMS}$ for our three most precisely-timed pulsars---\psre, \psrh, and \psrl---which have $\delta t_\mathrm{RMS} < 10$\,\us.
Also, without fitting for DMX, the timing fit for \PSRl was somewhat poor, reduced $\chi^2 \sim 3$.
Several additional parameters, such as $\ddot{\nu}$, $\dot{\nu}_\mathrm{b}$, $\dot{x}$, $\dot{e}_1$, and $\dot{e}_2$, appeared to be marginally significant, but their inclusion did not greatly improve $\chi_\mathrm{red}^2$.
PSRs \psre and \psrh also had $\chi_\mathrm{red}^2 \sim 2$ before DMX was introduced.

We split the observations of these pulsars into 6.5-day epochs and fit for one DMX parameter per epoch. 
We disabled the solar wind model in \tempotwo by setting the solar wind density at 1\,AU distance from the Sun to zero, so that all DM variations were modeled by DMX, including those due to the solar wind.
Certain observations of \PSRl presented in \citet{slr+14} have only single-frequency TOAs.
In some cases, the corresponding raw data are now missing, meaning we could not re-create the TOAs with retained frequency information.
For epochs containing only such TOAs, we fixed the value of the piecewise constant DMX function to zero.
In each case, the fit was improved after adding DMX, and additional orbital parameters were rendered insignificant for \PSRl.

\section{Results}\label{sec:results}
\begin{figure}
\centering
\includegraphics[width=0.48\textwidth]{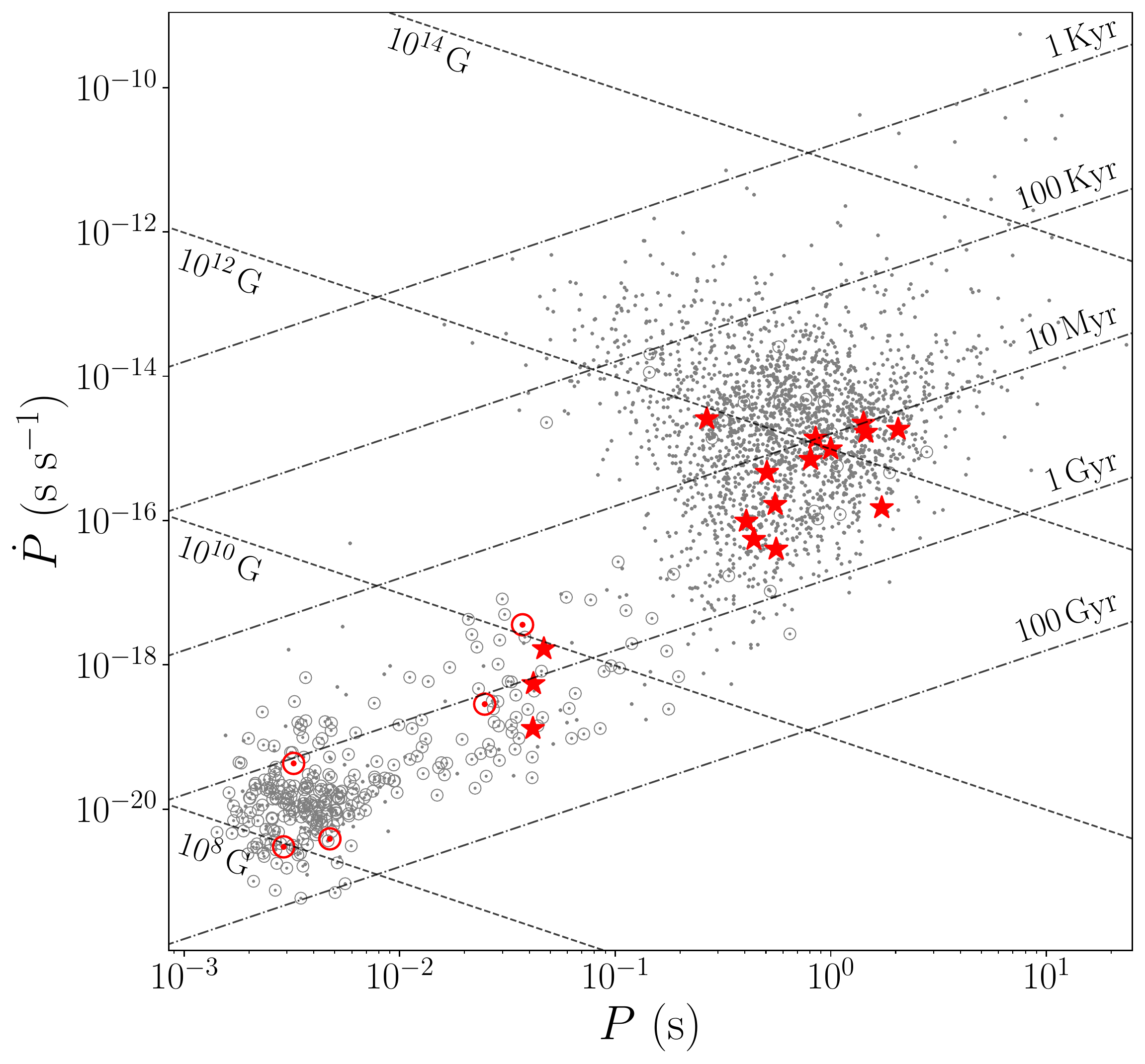}
\centering
\caption{
The pulsars presented in this work are plotted as red stars (or as dots surrounded by circles, if in binary systems) on this $P$--\pdot diagram, with values for the rest taken from the ATNF pulsar catalog. 
Lines of constant minimum surface magnetic field $B_\mathrm{surf}$ (in Gauss; dashed lines) and characteristic age $\tau_\mathrm{c}$ (dotted-dashed lines) are also shown.
}
\label{fig:ppdot}
\end{figure}

Each pulsar's spin frequency $\nu$ and frequency derivative $\dot{\nu}$ are given in Table~\ref{tab:rot-timing} along with general information about each timing solution.
We estimated distances to each of the pulsars based on their DMs, using the NE2001 \citep{ne2001:2002,ne2001:2003} and YMW16 \citep{ymw+17} Galactic free electron density models, reporting the distances given by each model.
These are reported, along with positions in Right Ascension (R.A.) and Declination (decl.) and in Galactic coordinates, in Table~\ref{tab:pos-dm}. 
Derived quantities are listed in Table~\ref{tab:derived}.
We show a $P$--\pdot diagram of our pulsars, as well as all pulsars in the ATNF pulsar catalog \citep{mht+05}, in Figure~\ref{fig:ppdot}.

\input{rot-timing.tex}\label{tab:rot-timing}
\input{pos-dm.tex}\label{tab:pos-dm}
\input{derived.tex}\label{tab:derived}

\begin{figure*}
\centering
\includegraphics[width=1.0\textwidth]{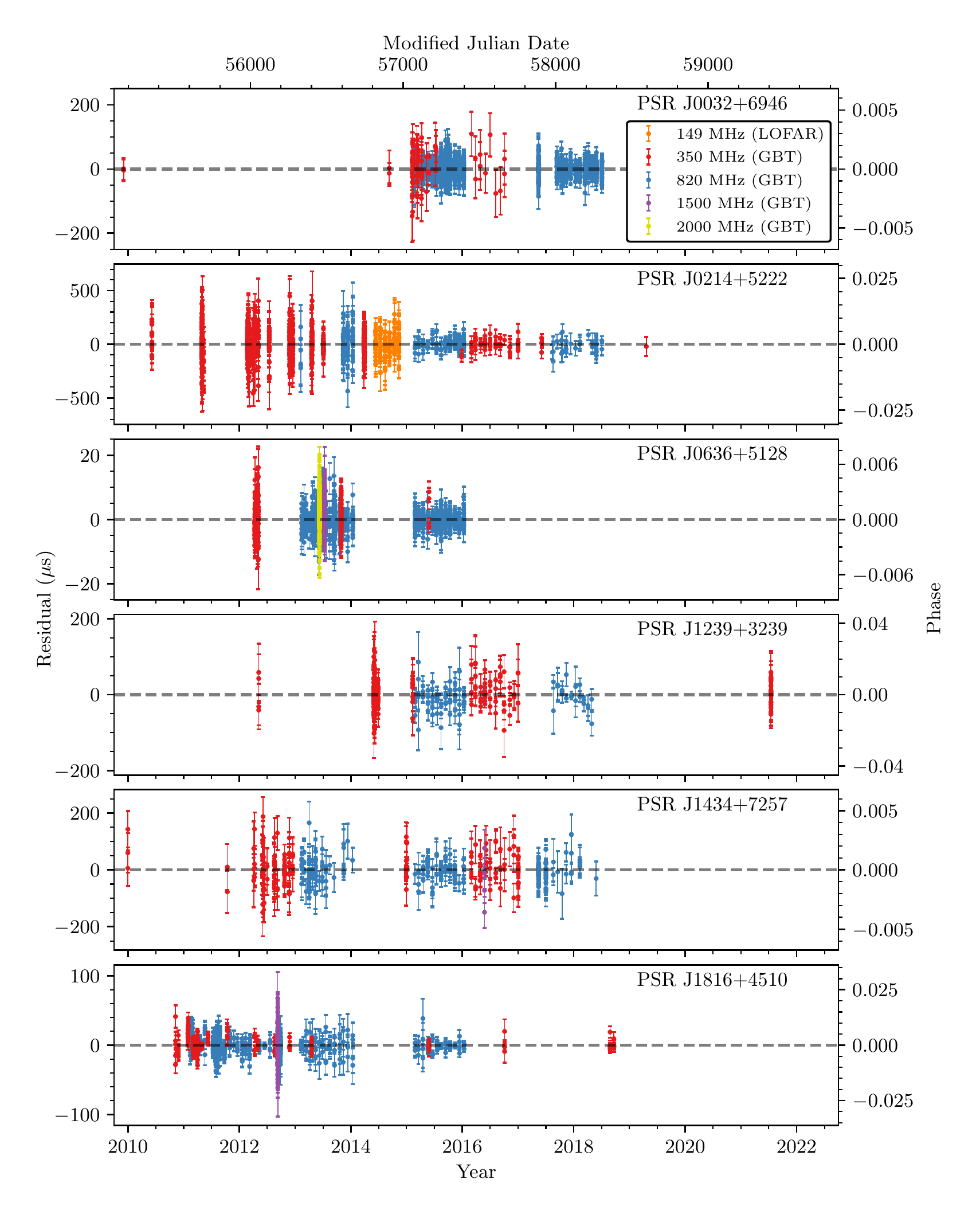}
\centering
\caption{
Timing residuals for six of the pulsars in this analysis. 
Residuals at 149\,MHz from LOFAR observations are plotted in orange, while 350, 820, 1500, and 
2000\,MHz 
residuals from GBT observations are plotted in red, blue, purple, and yellow, respectively.
The dashed gray lines correspond to a residual of zero.
Error bars are the 1-$\sigma$ uncertainties in each TOA.}
\label{fig:res1}
\end{figure*}

\begin{figure*}
\centering
\includegraphics[width=1.0\textwidth]{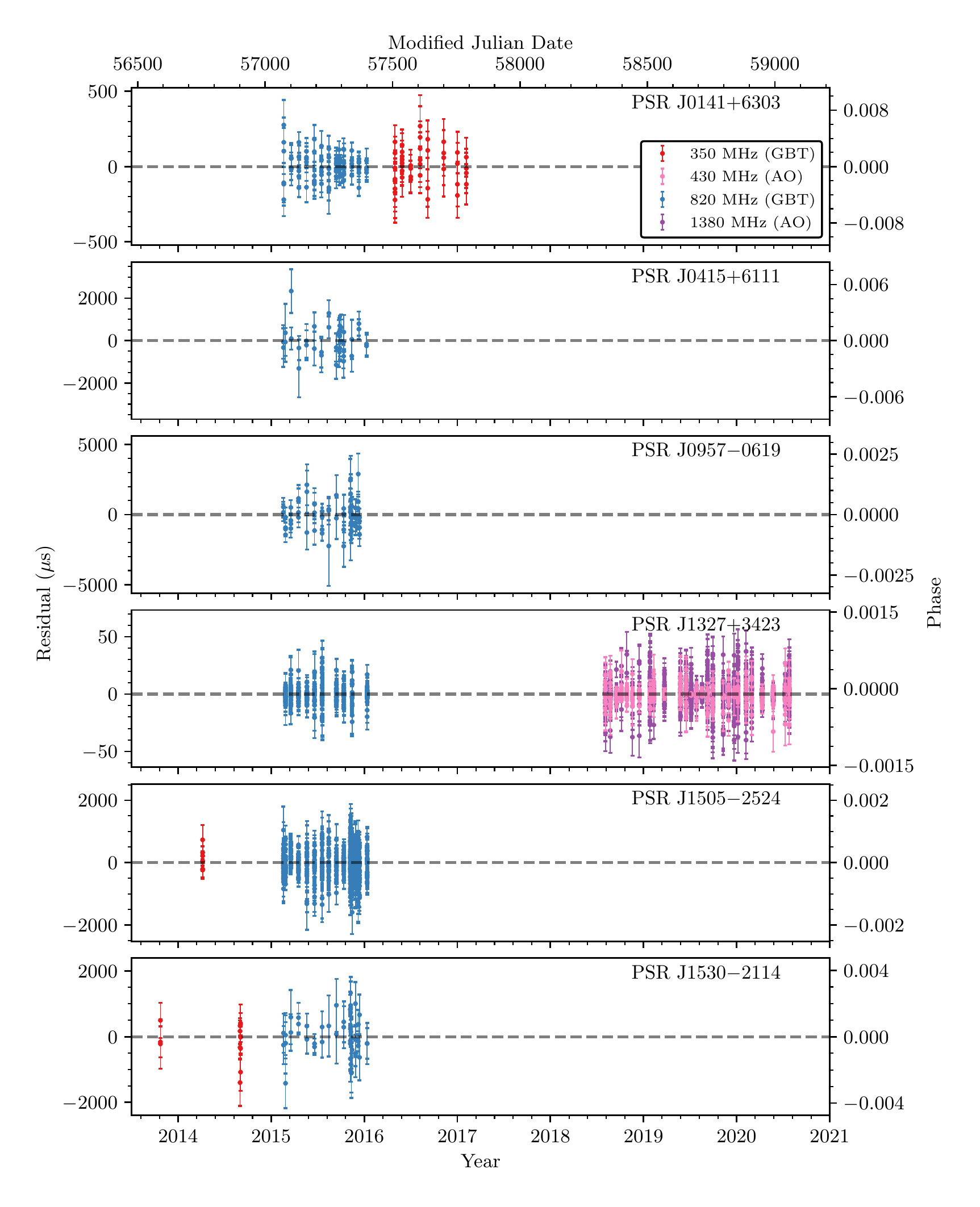}
\centering
\caption{
Timing residuals for six  pulsars, with 430 and 1380\,MHz residuals from NANOGrav AO observations plotted in pink and purple, respectively, and 350 and 
820\,MHz 
residuals from GBT observations  plotted in red and blue, respectively.
The dashed gray lines correspond to a residual of zero.
Error bars are the 1-$\sigma$ uncertainties in each TOA.}
\label{fig:res2}
\end{figure*}

\begin{figure*}
\centering
\includegraphics[width=1.0\textwidth]{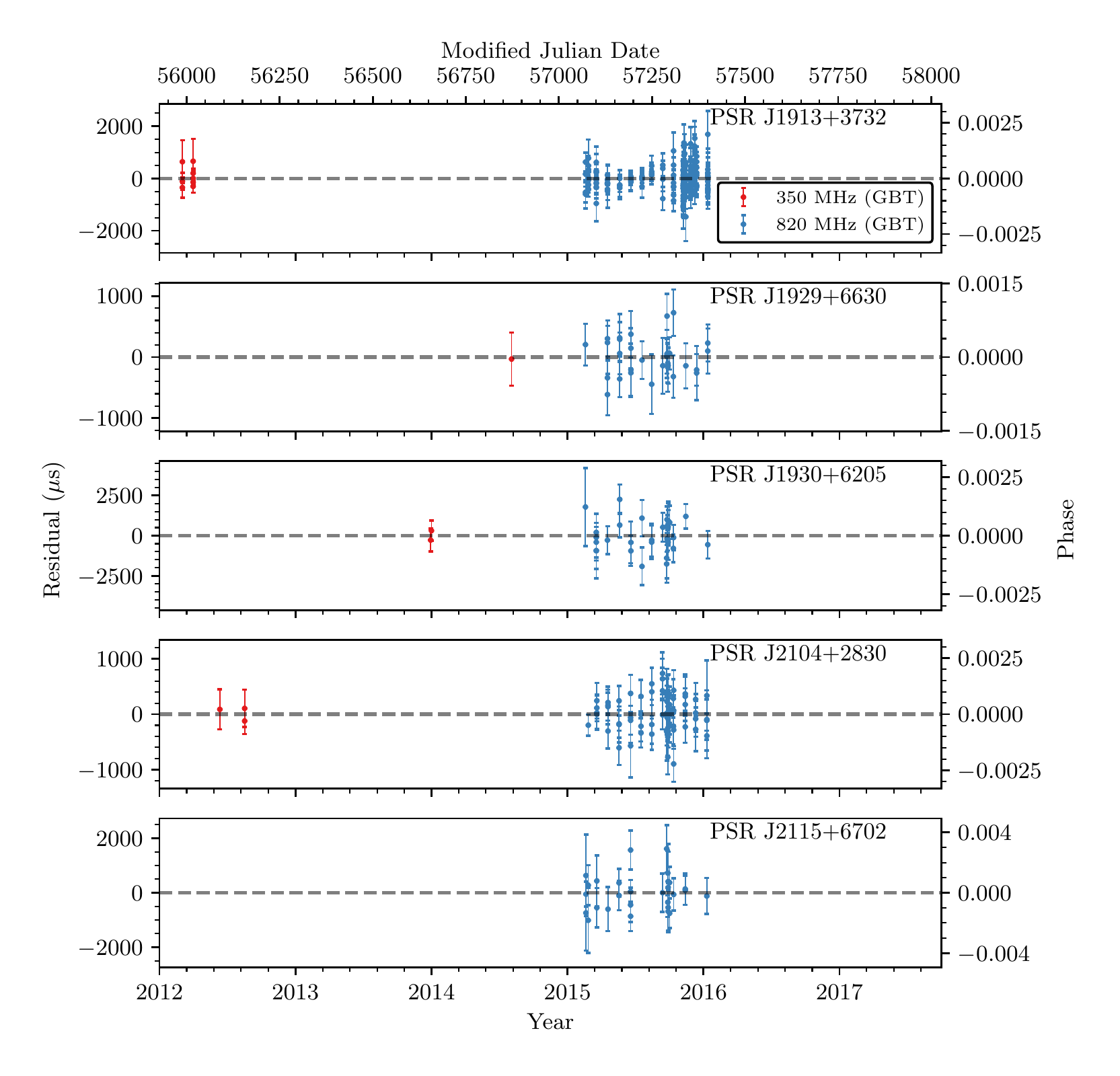}
\centering
\caption{
Timing residuals for five pulsars, with 350 and 
820\,MHz 
residuals from GBT observations plotted in red and blue, respectively.
The dashed gray lines correspond to a residual of zero.
Error bars are the 1-$\sigma$ uncertainties in each TOA.}
\label{fig:res3}
\end{figure*}

\begin{figure*}
\centering
\includegraphics[width=1.0\textwidth]{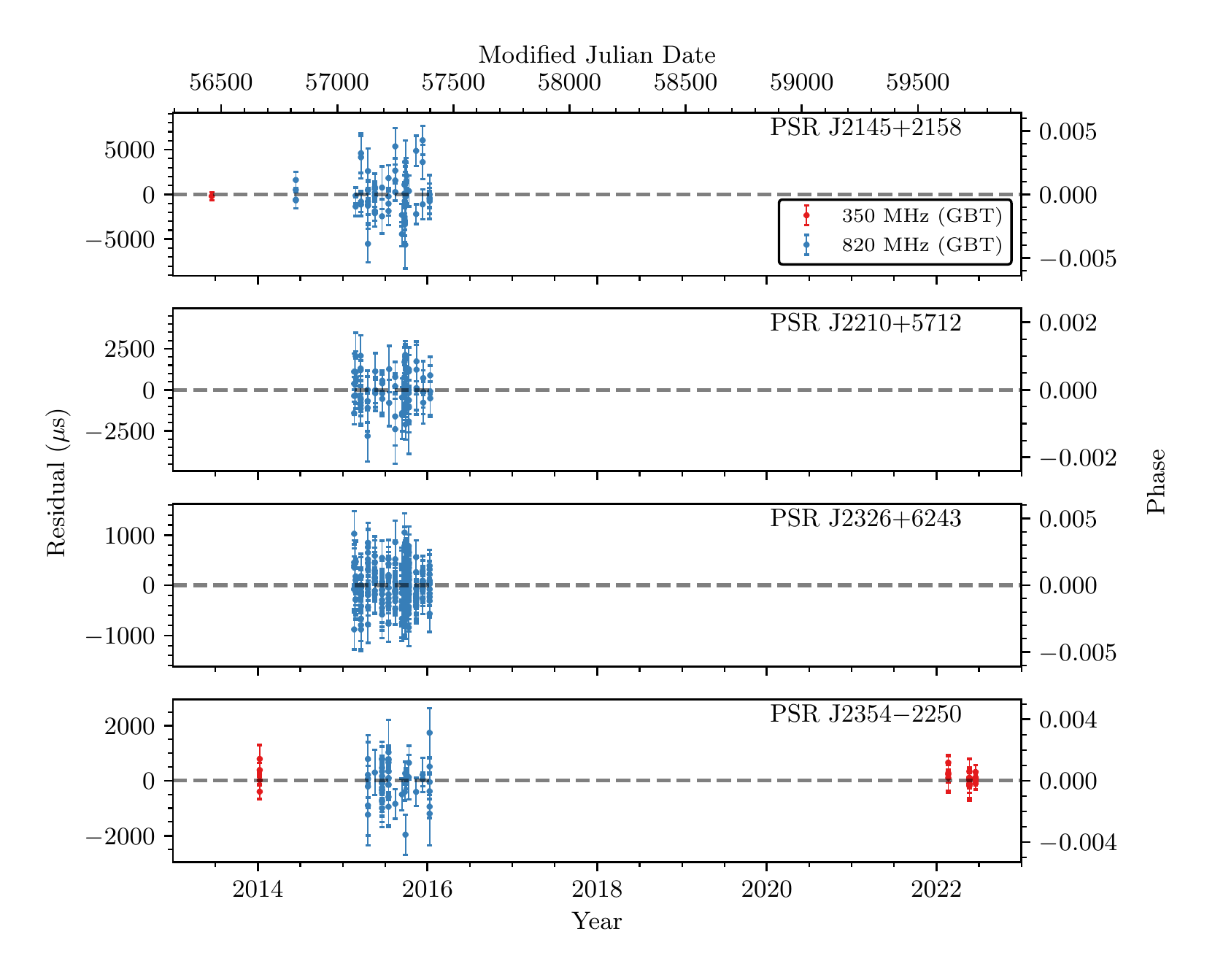}
\centering
\caption{
Timing residuals for the final four pulsars, with 350 and 
820\,MHz 
residuals from GBT observations plotted in red and blue, respectively.
The dashed gray lines correspond to a residual of zero.
Error bars are the 1-$\sigma$ uncertainties in each TOA.}
\label{fig:res4}
\end{figure*}

\begin{figure}
\centering
\includegraphics[width=0.49\textwidth]{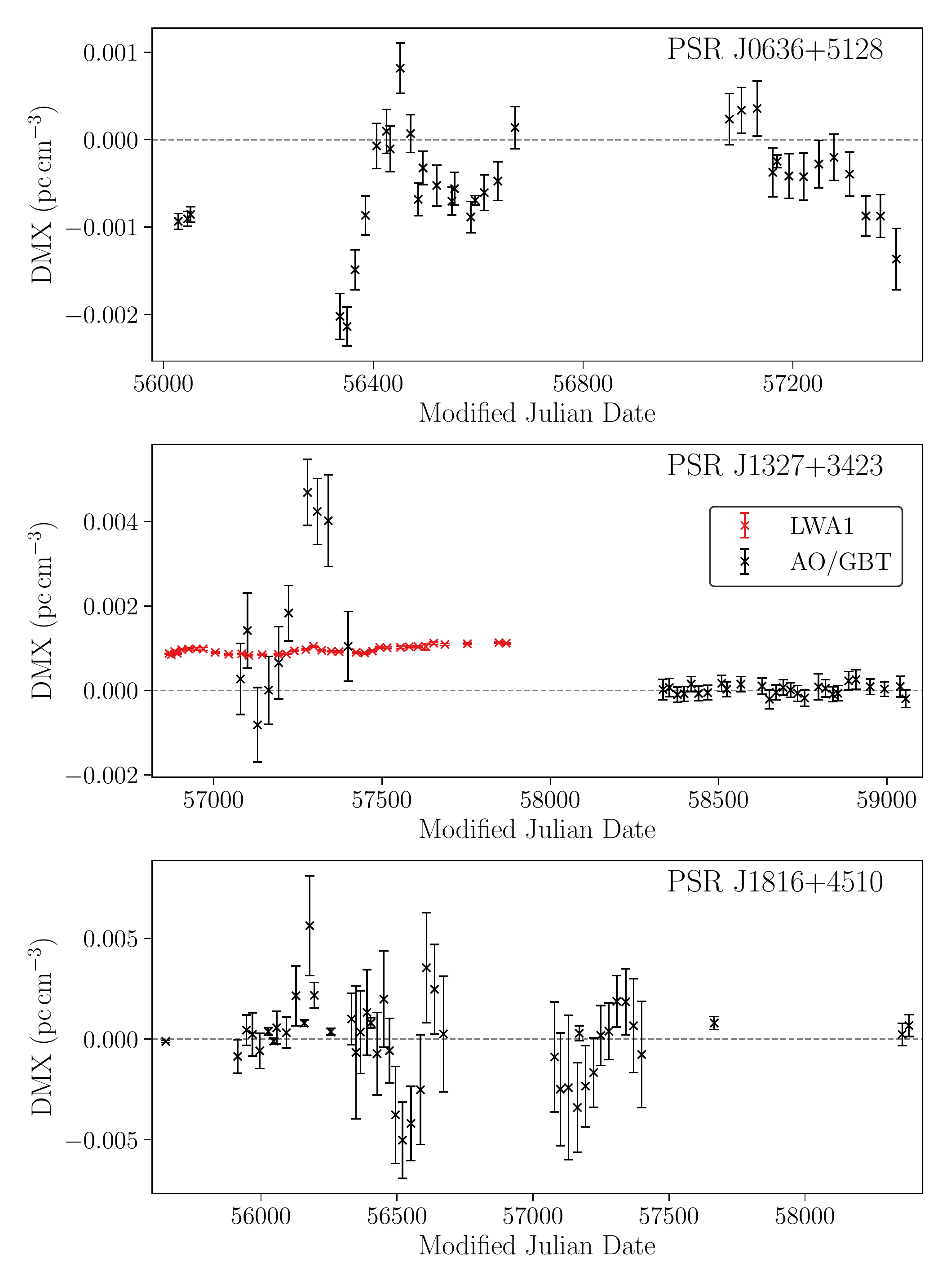}
\centering
\caption{
DMX timeseries for PSRs \psre, \psrh, and \psrl, showing the offset from a fiducial DM value at each epoch. The dashed gray line in each plot corresponds to an offset of zero.
The plot for \PSRh shows DMX in 
red corresponding to LWA1 observations; the points in black are from the GBT (before MJD 58000) and AO (after MJD 58000). 
Epochs for which we were unable to fit for DMX (described in Section~\ref{sec:dmx}) are not shown. 
Error bars correspond to the 1-$\sigma$ uncertainty in DMX reported by \tempotwo.}
\label{fig:DMX}
\end{figure}

\begin{figure*}
\centering
\includegraphics[width=1.0\textwidth]{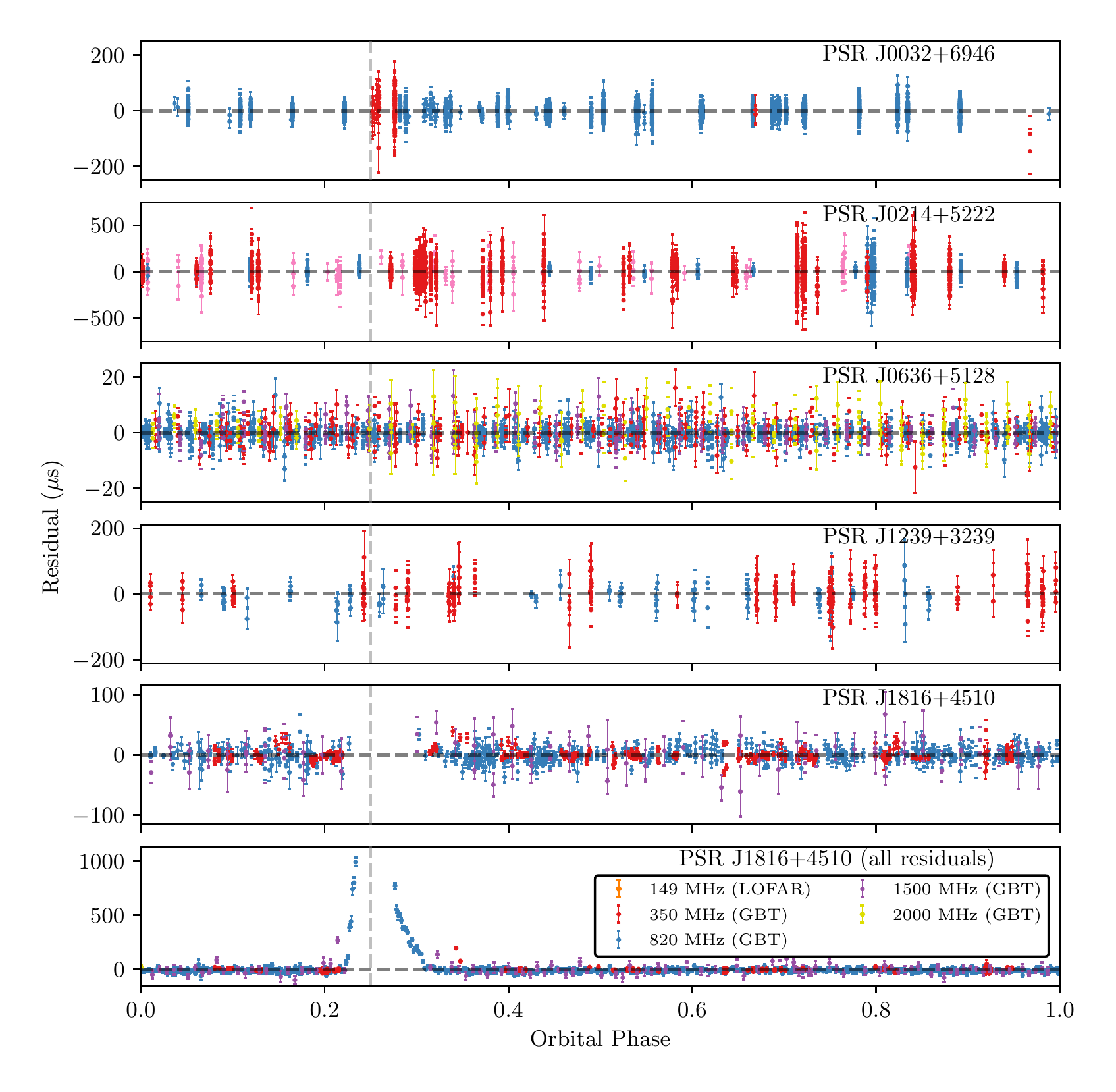}
\centering
\caption{
Timing residuals for the five pulsars in binary systems are plotted vs.~orbital phase (see Equation~\ref{eq:phase}, but one orbit = 1.0 instead of $2\pi$ radians). 
The gray dashed vertical line at $\phi = 0.25$ denotes superior conjunction, and the horizontal dashed lines correspond to a residual of zero.
The second set of residuals from the bottom are the residuals for \PSRl which contributed to our timing solution.
The bottom-most set of residuals are all of our residuals for \PSRl, including those which exhibit additional timing delays before and after eclipse. 
In the latter set, TOAs from the full-orbit observations of \PSRl were created with shorter sub-integrations so that the behavior near eclipse can be seen more easily. The 350, 820, 1500, and 
2000\,MHz 
residuals from GBT observations are plotted in red, blue, purple, and yellow, respectively.
Error bars are the 1-$\sigma$ uncertainties in each TOA.}
\label{fig:res_orb}
\end{figure*}

\vspace{-26.5mm}
Figures \ref{fig:res1}, \ref{fig:res2}, \ref{fig:res3}, and \ref{fig:res4} show each pulsar's final set of timing residuals.
DMX timeseries are shown in Figure~\ref{fig:DMX}.
PSRs \psra, \psrc, \psre, \psrg, and \psrl are in binary systems.
Timing residuals for these pulsars are plotted against orbital phase in Figure~\ref{fig:res_orb}, and we list pulsars' best-fit orbital parameters in Table~\ref{tab:binaries}.

\input{binaries.tex}\label{tab:binaries}

\vspace{-9mm}
We measured proper motions in R.A. ($\mu_\alpha = \dot{\alpha}\cos\delta$, where $\alpha = \mathrm{R.A.}$ and $\delta = \mathrm{decl.}$) and decl.~($\mu_\delta = \dot{\delta}$) for five pulsars with several years of timing data: PSRs \psrc (only $\mu_\alpha$ is significant), \psre, \psrh, \psri, and \psrl. 
Using total proper motions and DM-derived distances, we calculated transverse velocities $v_t$ for these pulsars. 
This allows the determination of the apparent rate of spindown due to the relative transverse motion between the pulsar and solar system barycenter, known as the Shklovskii effect \citep{sis+70}, which we write as \ps.
Using the same method laid out in \citet{spp+23}, based on \citet{gfg+21}, we also calculate a correction \pg due to the pulsar's acceleration in the Galactic potential, using the most recent value for the distance between the Sun and the Galactic center, $R_0 = 8.275(34)$\,kpc \citep{hf+04}, and for the circular velocity of the Sun around the Galactic center, $\Phi_0 = 240.5(4.1)$\,km\,s$^{-1}$ \citep{gravity+21}.

Together, these corrections give us each pulsar's intrinsic rate of spindown, $\dot{P}_\mathrm{int} = \dot{P} - \dot{P}_\mathrm{S} - \dot{P}_\mathrm{G}$, which we then use to re-calculate other derived parameters which depend on $\dot{P}$.
We list these corrected quantities, along with proper motions and transverse velocities, in Table~\ref{tab:pms}.
Each pulsar has two sets of calculated velocities and corrections, each assuming either the NE2001 or YMW16 DM distance.

\input{pms.tex}\label{tab:pms}
\vspace{-9mm}
Parameters only measured for a few pulsars include a frequency-dependent (FD) parameter which accounts for radio-frequency-dependent profile evolution \citep{abc+15} and timing parallax $\varpi$.
These are given in Table~\ref{tab:misc}, along with a value of $\varpi$ corrected for Lutz-Kelker bias \citep{vlm+10}, and a corresponding parallax distance.

\input{misc.tex}\label{tab:misc}

\subsection{Disrupted Recycled Pulsars}\label{sec:DRPs}
With $P$~$\sim$~40\,ms and $B_\mathrm{surf} < 3 \times 10^{10}$\;G, and no evidence of a binary companion, PSRs \psrb and \psrh are new DRPs. 
\PSRi is also a DRP; an initial timing solution for that pulsar was published in \citet{slr+14}, and we have now measured its proper motion, which is presented in Table~\ref{tab:pms}. 
We also measured proper motion for \PSRh, along with timing parallax: $\varpi = 4(1)$\,mas. 
We corrected this parallax measurement for Lutz-Kelker bias using the online tool\footnote{\url{http://psrpop.phys.wvu.edu/LKbias/}} provided by \citet{vlm+10}, and obtained a corrected value of 1.1$^{+1.1}_{-0.5}$\;mas. This implies a distance of 0.9$\mathrm{^{+0.8}_{-0.5}}$\;kpc, which further implies $v_\mathrm{T} = 4(3)$\,km\,s$^{-1}$.
This distance is higher than, but marginally consistent with, the DM-derived distances (0.5 and 0.3\,kpc using the NE2001 and YMW16 electron density models, respectively).
Tension between parallax and DM distances is not unusual; \citet{dgb+19} found greater than factor-of-three discrepancies between DM distances and parallax distances measured using Very Long Baseline Interferometry.

It has been hypothesized that due to kicks received by their disrupting supernovae, DRPs have high space velocities, and may lie further off the Galactic plane than, e.g., DNS binaries \citep{lma+04}. 
\citet{kmk+18} found this was the case, with DRPs' median and mean $z$-height off the plane being 385 and $580 \pm 160$\,pc, respectively, versus 200 and $300 \pm 100$\,pc for DNS systems.
Based on these pulsars' DM-derived distances, their $z$-heights are consistent with those of other DRPs: $z = 120$ -- 580\,pc (for \PSRb; given as a range between values corresponding to the NE2001 and YMW16 electron density models), 290 -- 490\,pc (for \PSRh), and 470 -- 670\,pc (for \PSRi).
However, our measured transverse velocities for PSRs \psrh and \psri are not particularly large: 15 -- 21 and 30 -- 40\,km\,s$^{-1}$, respectively.

\vspace{3mm}
\subsection{The Low-Frequency DM of \PSRh}\label{sec:chromaticdm}
As described in Section~\ref{sec:observations}, we observed \PSRh with LWA1 at very low radio frequencies, 26 -- 88\,MHz.
The resulting TOAs have large uncertainties compared to those resulting from GBT or AO observations, $\sigma_\mathrm{TOA} \sim 100$\,\us, leading us to disregard them for our regular timing analysis
(this is why LWA1 TOAs are not represented in the timing residuals plotted in Figure~\ref{fig:res2}).

However, precise measurements of DM are made possible by low-frequency observations.
We produced a DMX timeseries from these data, holding all other timing parameters fixed.
This is plotted in Figure~\ref{fig:DMX}, along with the DMX timeseries corresponding to AO and GBT observations.
The average DM we measure from LWA1 observations is $\approx$0.001\,\dmu higher than the DM we measure from AO observations, which also have relatively precise DM measurements, thanks to the nearly-simultaneous observations at 430 and 1380\,MHz.

\begin{figure}
\centering
\includegraphics[width=0.49\textwidth]{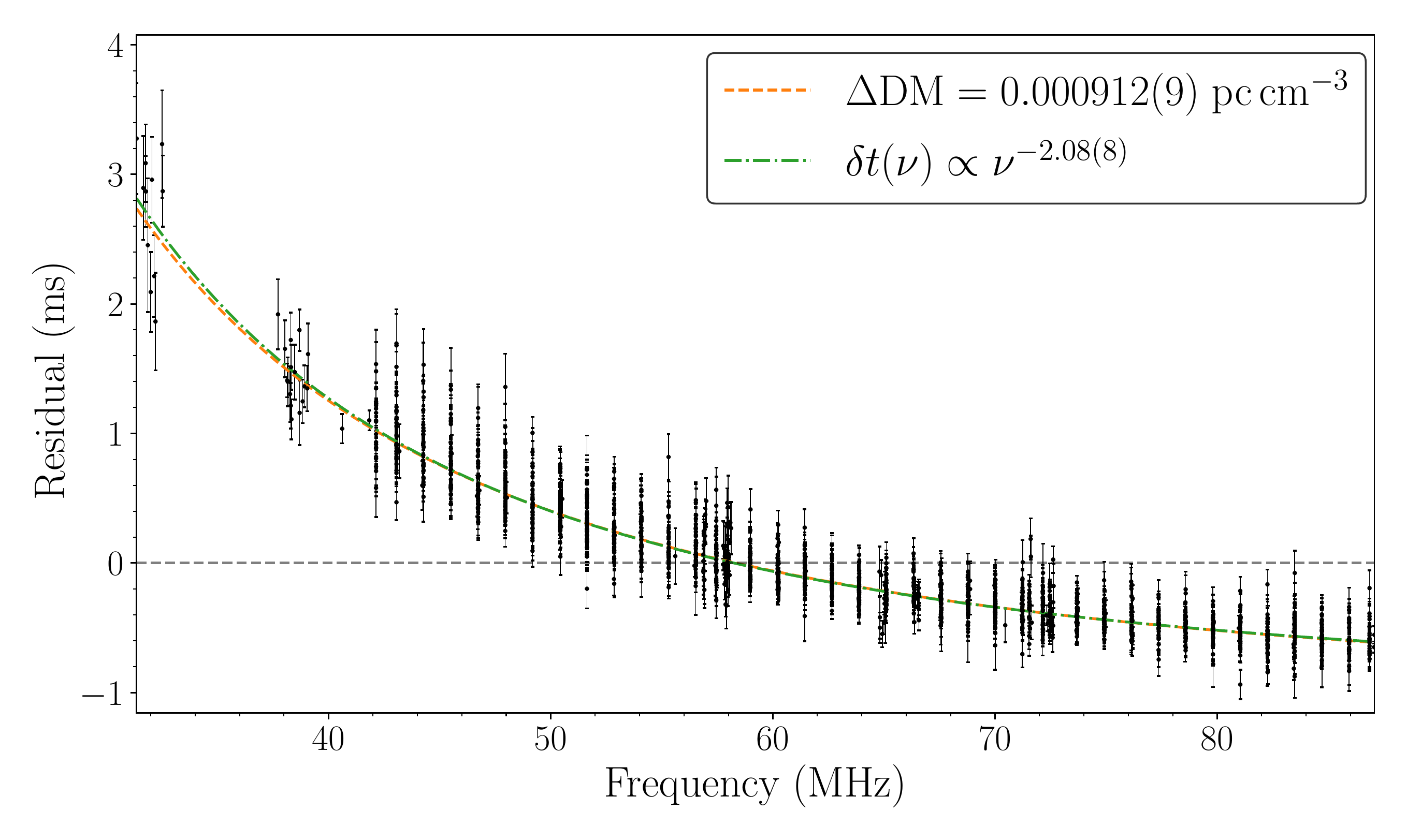}
\centering
\caption{Timing residuals for \PSRh which compare LWA1 TOAs to our AO/GBT timing solution are shown, plotted against radio frequency. 
The dashed gray line shows $\delta t = 0$.
The curved dashed line shows our fit to the residuals assuming a $\nu^{-2}$ dependence, corresponding to a DM 0.000912(9)\,\dmu higher than the fiducial DM used in our timing model (modulo a constant offset).
The dotted-dashed line shows a fit which does not make that assumption.
The dependence is consistent with a dispersive delay.
}
\label{fig:LWAres}
\end{figure}

In Figure~\ref{fig:LWAres}, we show LWA1 timing residuals versus frequency for \PSRh, with respect to the timing solution reached with AO/GBT data.
Assuming the strong frequency-dependent delay seen in the residuals is caused by an increase $\Delta$DM in the pulsar's DM, the delay can be described by
\begin{equation}\label{eq:ddm}
    \delta t(\nu) = 4.15 \times 10^{6}\;\ms\bigg(\frac{\Delta\mathrm{DM}}{\mathrm{pc\,cm}^{-3}}\bigg)\bigg(\frac{\nu}{\MHz}\bigg)^{-2}.
\end{equation}
We performed a least-squares fit to the residuals, recovering $\Delta\mathrm{DM} = 0.000912(9)$\,\dmu.
We also performed a fit without assuming a $\nu^{-2}$ dependence, finding that $\delta t \propto \nu^{-2.08(8)}$, consistent with dispersion.

It is possible that profile frequency evolution could cause a signature in the residuals which might lead to a mis-estimated DM.
The LWA1 TOAs were created using separate (aligned) standard profiles for each of the four center frequencies: 35.1, 49.8, 64.5, and 79.2\,MHz, which should account for nearly all of the frequency evolution in the band.
Inspecting the relative shapes of the LWA1 pulse profiles, the width of the profile broadens towards lower frequency, but we do not see any other evolution in the shape of the profile within the LWA1 band.
In fact, as can be seen from Figure~\ref{fig:profs2}, this pulsar's profile shape is remarkably stable throughout the entire range of observed frequencies, from 35\,MHz to over 1.38\,GHz.

It is also possible that this effect is due to a frequency-dependent DM.
This occurs because radio waves emitted by the pulsar are scattered by the interstellar medium, taking different paths between the pulsar and the Earth.
Because the transverse extent of the path lengths varies with frequency, observations at different frequencies sample different regions of the ISM.
$\sim$AU-scale electron density fluctuations can then cause DM to vary with frequency \citep{css+16}.

We see strong scintillation in our 820 and 1380\,MHz observations of this pulsar.
Our observations are not long enough to resolve the scintles in time, but the scintillation bandwidth is $\sim$75\,MHz.
We did not perform a complete scintillation analysis, so this is a rough estimate.
Using equation 12 in \citet{css+16}, we can roughly calculate the expected two-frequency DM difference, assuming a Kolmogorov medium and a thin screen.
If we take our two frequencies to be $\nu \sim 0.9$\,GHz and $\nu' = 57$\,MHz, then we can expect an rms DM difference of order $0.009\,\mathrm{pc\,cm}^{-3}$, which is just within an order of magnitude of our measured DM difference.

\subsection{Mildly Recycled Wide Binary Pulsars}\label{sec:wide}
PSRs \psra and \psrc belong to a small group of mildly recycled pulsars in wide binary systems, with orbital periods $P_\mathrm{b} > 200$\,days.
There are four other such pulsars: J0407+1607 \citep{lxf+05}, J1840$-$0643 \citep{kek+13}, J2016+1948 \citep{naf+03}, and J2204+2700 \citep{mgf+19}. 
The rotational properties of these pulsars imply histories of recycling.
This likely occurred when their companion stars entered the giant phase, allowing accretion to occur despite the wide orbital separations.
However, perhaps because of the companions' higher masses, and hence shorter evolutionary timescales, recycling ended before these pulsars could become MSPs \citep{tlp+15}.

Both pulsars were first published as discoveries in \citet{slr+14}, but a fully-coherent timing solution did not yet exist for \PSRa; its binary motion was noticed later.
It has a spin period of 37\,ms and is in a 523-day, relatively circular ($e \simeq 0.0005$) orbit with a $0.42$\,\Msun$ < M_\mathrm{c} < 1.2$\,\Msun companion\footnote{
An upper bound on $M_\mathrm{c}$ can be placed at the 90\% confidence level by taking $i = 26\arcdeg$, as randomly-distributed inclination angles will fall under that value only $\approx$10\% of the time \citep{lk+04}.
}, likely a CO-core white dwarf \citep[WD;][]{ttm+11}.
This range of allowed companion masses is consistent with \citet{tsg+99}, who predict the mass of a WD in a $\sim$500-day orbit with a recycled pulsar to be $M_\mathrm{c} \sim 0.4 - 0.45$\,\Msun.

Our updated timing solution for \PSRc ($P_\mathrm{b} \simeq 512$\,days, $e \simeq 0.005$ and the same $M_\mathrm{c}$ range) is consistent with the \citet{slr+14} solution, except for the introduction of two new parameters: $\mu_\alpha = 9(1)\,\mathrm{mas}\,\yr^{-1}$ ($\mu_\delta$ is not significant; the best-fit value is 2(2)\,mas\,yr$^{-1}$ according to \texttt{PINT}), and a secular change in the projected semimajor axis, $\dot{x}_\mathrm{obs} = -2.1(6) \times 10^{-13}\;\s\,\s^{-1}$.
Secular changes in $x$ have several potential causes:
\begin{equation}
    \dot{x}_\mathrm{obs} = \dot{x}_\mathrm{GW} + \dot{x}_\mathrm{PM} + \dot{x}_\mathrm{kin} + \dot{x}_{\dot{m}} + \dot{x}_\mathrm{prec} + \dot{x}_\mathrm{planet},
\end{equation}
where $\dot{x}_\mathrm{GW}$ is the contribution from gravitational-wave damping, $\dot{x}_\mathrm{PM}$ is caused by the proper motion of the system, $\dot{x}_\mathrm{kin} = \dot{x}_\mathrm{G} + \dot{x}_\mathrm{S}$ is due to the two kinematic effects described in Section~\ref{sec:results} (Galactic acceleration, $\dot{x}_\mathrm{G}$, and the Shklovskii effect, $\dot{x}_\mathrm{S}$), $\dot{x}_{\dot{m}}$ is due to mass loss, $\dot{x}_\mathrm{prec}$ is caused by precession (including geodetic precession of the pulsar's spin axis and orbital precession caused by spin-orbit coupling), and $\dot{x}_\mathrm{planet}$ is caused by the deformation of the orbit due to a third body, such as a planet \citep{lk+04}.

We do not expect $\dot{x}_{\dot{m}}$ or $\dot{x}_\mathrm{prec}$ to be significant for this pulsar, due to its wide orbit.
The contribution from GW damping is $\dot{x}_\mathrm{GW} \approx 10^{-26}\;\s\,\s^{-1}$.
For $\dot{x}_\mathrm{kin}$, we can use the values for \pg and \ps reported Table~\ref{tab:pms} to relate the change in apparent \pdot to the variation in $x$ which is caused by the same underlying effects:
\begin{equation}\label{xdotkin}
    \dot{x}_\mathrm{kin} = -(\dot{P}_\mathrm{G} + \dot{P}_\mathrm{S})(x/P) \approx 3.5 \times 10^{-17}\;\s\,\s^{-1}.
\end{equation}

We cannot rule out the presence of a planetary object. 
However, to produce the observed $\dot{x}$, a planet would either need to be placed rather conveniently, have an extremely wide orbit ($r \sim 400$\,AU) or both \citep[see Equation 8.85 in][]{lk+04}.
Assuming no such object exists, the only effect which can explain our observed $\dot{x}$ is a gradual change in the inclination angle of the binary system due to its proper motion.

Measuring this secular variation in $x$, together with that of $\omega$, can allow one to determine $i$ and the longitude of ascending node, $\Omega$ \citep{ksm+96}.
Assuming our measurement is entirely due to proper motion, and that $\mu_\delta = 0$, we can use Equation 11 in \citet{ksm+96} to say the following:
\begin{equation}\label{eq:kopeikin}
    \cot i\sin\Omega = 0.9(0.3).
\end{equation}
With such high relative uncertainty, we cannot confidently rule out any range of values for $i$.
To 68\% confidence, $i < 59\arcdeg$.

\subsection{PSR J1239+3239}\label{sec:J1239}
\PSRg is an MSP with a spin period of 4.7\,ms and is in a 4-day orbit with a low-mass companion. Pulsar timing limits the mass of the companion to the range $0.13 < M_\mathrm{c} < 0.31$.
This is consistent with the prediction of \citet{tsg+99} for the mass of a WD with this orbital period: $M_\mathrm{c} \sim 0.22$\,\Msun.

This pulsar has $\delta t_\mathrm{RMS} = 21.5$\,\us.
Furthermore, its wide pulse profile broadens at higher frequencies.
Due to these factors, this MSP is unfortunately not suitable for inclusion in PTAs.

\vspace{3mm}
\subsection{PSR J0636+5128}\label{sec:J0636}
\PSRe (originally known as PSR J0636+5129) is a black widow MSP with a spin period of 2.8\,ms in a tight, 
1.62\,hr
orbit with an extremely-light companion: 
$0.007\,\mathrm{M}_\odot < M_\mathrm{c} < 0.016\,\mathrm{M}_\odot$.
Assuming $i=60\arcdeg$ and a pulsar mass of $M_\mathrm{p} = 1.4$\,\Msun, $M_\mathrm{c} = 8.5\,\mathrm{M_{Jupiter}}$. 
\citet{slr+14} proposed this companion as a possible ``diamond planet'' \citep[e.g.,][]{bbb+11,ngc+13}.

We have measured two significant orbital frequency derivatives (see Table~\ref{tab:misc}); the corresponding derivative of orbital period, $\dot{P}_\mathrm{b} = 6(1) \times 10^{-12}$, is of the wrong sign and two orders of magnitude too high to be explained by orbital decay from GW emission. \citet{ksk+18} found that kinematic effects due to motion relative to the solar system barycenter also could not explain the measured $\dot{P}_\mathrm{b}$.
Like in other black widow systems, these orbital frequency derivatives can be explained by tidal 
and wind effects \citep{as+94,cwc+21}.

While fitting for several orbital frequency derivatives is often necessary for black widow systems, this has been shown not to reduce PTAs' sensitivity to GWs \citep{brd+15}.
Furthermore, \PSRe does not have radio eclipses, and its timing residuals do not suggest the presence of ionized material in its orbit. 
Consequently, it has been observed by both NANOGrav and the EPTA \citep{aab+21,dcl+16}.

\citet{slr+14} reported a timing parallax of $\varpi = 4.9(6)$\,mas for this pulsar; this parameter is not part of our updated solution.
With our timing data, \tempotwo gives a best-fit value of $\varpi = 1.8(1.5)$\,mas, which despite not being significant, is consistent with the most recently-published NANOGrav timing solution, which has $\varpi = 1.4(2)$\,mas \citep{aab+21}. 
This change is likely due to our longer data span and use of the DMX model.
With a shorter data span, DM variations may have been subsumed into the parallax fit in the \citet{slr+14} solution \citep{ksk+18}.

\subsection{PSR J1816+4510}\label{sec:J1816}

\PSRl has a spin period of 3.2\,ms and is in an 8.66\,hr orbit with a $0.16\,\mathrm{M}_\odot < M_\mathrm{c} < 0.41\,$\Msun companion, consistent with the characteristics of redback systems \citep{rms+11}.
See \citet{ksr+12} and \citet{slr+14} for details on this pulsar's discovery and initial timing solution, which we are updating in this work.
Our updated parameters are largely consistent with the \citet{slr+14} solution, with the only significant change being a reduction in $\mu_\alpha$ from 5.3(8) to 2(1)\,\pmu.

\PSRl exhibits regular, short-duration eclipses, and the TOAs immediately before and after the eclipse exhibit delays up to 800\,\us due to excess material in the orbit.
The delays sharply increase at ingress, and slowly fade to normal post-eclipse over a period nearly as long as the eclipse itself (see the bottom plot of Figure~\ref{fig:res_orb}).
The spectrum of \PSRl's companion star is similar to He WDs, but it has the high metallicity and low surface gravity suggestive of a larger, possibly non-degenerate star \citep{kdk+13}. 
It may be that the companion is a bloated proto-He WD which has yet to reach the cooling track \citet{itl+14}.

\citet{pbs+20} studied the eclipses of \PSRl at 149 and 650\,MHz. They showed that eclipses last longer at lower frequencies, lasting for 12.5\% of the orbit at 149\,MHz compared to 5\% at 650\,MHz. Contrary to \citet{slr+14}, they differentiate between the true eclipse (where the pulsar emission disappears entirely) and a ``smearing'' phase, where additional delays are observed in the TOAs.

To better characterize the eclipse duration and investigate the additional delays during ingress and egress, we producing a single TOA for each 53\,s of data (0.17\% of the orbit) for our full-orbit GBT observation at 820\,MHz, and 4\,min (0.77\%) for the corresponding 1500\,MHz observation. 
At 820\,MHz, the eclipse began at an orbital phase (in units of rotations; otherwise as in Equation~\ref{eq:phase}; superior conjunction occurs at $\phi = 0.25$.) of $\phi \approx 0.236$ and lasted until $\phi \approx 0.276$, for a duration of $\approx$ 4\% of the orbit, or $\approx$ 21 minutes.
At 1500\,MHz, eclipse appears to last for $\sim$ 5\% of the orbit.
However, S/N is low at this frequency, so our ability to differentiate between the ``eclipse'' and ``smearing'' phases may be limited.
Curiously, the 1500\,MHz TOA immediately before eclipse does not appear to have significant additional delay, but the TOA before that does have a large residual, $\delta t \approx 265$\,\us.
The two 1500\,MHz TOAs after eclipse behave similarly, with the first consistent with zero delay and the next being delayed by $\delta t \approx 135$\,\us.

We see some signs that the length of eclipse varies from orbit to orbit, especially at lower frequencies.
We have a lack of orbital phase coverage at 350\,MHz from $0.22 \lesssim \phi \lesssim 0.32$.
We see no excess delay in a 350\,MHz observation at $\phi \approx 0.32$, though delays do seem to be present in observations at $\phi \approx 0.35$, from 39\,\us in one observation to 200\,\us in another.
\citet{pbs+20} see similar variations in their 149\,MHz observations and suggest they are due to clumping in the tail of material extending from the companion star.

\PSRl is associated with the gamma-ray source 4FGL J1816.5+4510. Recently, \citet{cbb+23} reported the discovery of gamma-ray eclipses in several pulsars, including \PSRl. The existence of such eclipses requires that the companion star directly occult the pulsar, constraining $i$, and thus $M_\mathrm{p}$. 
The constraints they report are $i > 79.0\arcdeg$ and allowed ranges $1.64 \leq M_\mathrm{p} \leq 2.17~\mathrm{M}_\odot$ and $0.18 \leq M_\mathrm{c} \leq 0.22~\mathrm{M}_\odot$ for a Roche-lobe-filling companion, or $i > 82.6\arcdeg$ and $1.68 \leq M_\mathrm{p} \leq 2.11~\mathrm{M}_\odot$ for a Roche-lobe filling factor of 0.5.

We can use this constraint on $i$ to estimate the companion's size using the observed eclipse duration at 820\,MHz. 
Assuming $M_\mathrm{p} = 1.64$\,\Msun and $i = 90\arcdeg$, we calculate $a = 2.6$\,\Rsun.
The radius for the object which obscures the pulsar for 4\% of the orbit is therefore $R_\mathrm{c} = 0.33$\,\Rsun.
We can also estimate the companion's Roche lobe radius, $R_\mathrm{L} = 0.55$\,\Rsun. 
These values do not differ greatly in the range $80\arcdeg \leq i \leq 90\arcdeg$.
It is therefore clear that \PSRl's companion does not extend beyond its Roche lobe, aside from the aforementioned tail of ionized material, which extends for a distance $\sim$ the diameter of the companion.

\subsection{Nulling Pulsars}\label{sec:nullers}

\begin{figure}
\centering
\includegraphics[width=0.48\textwidth]{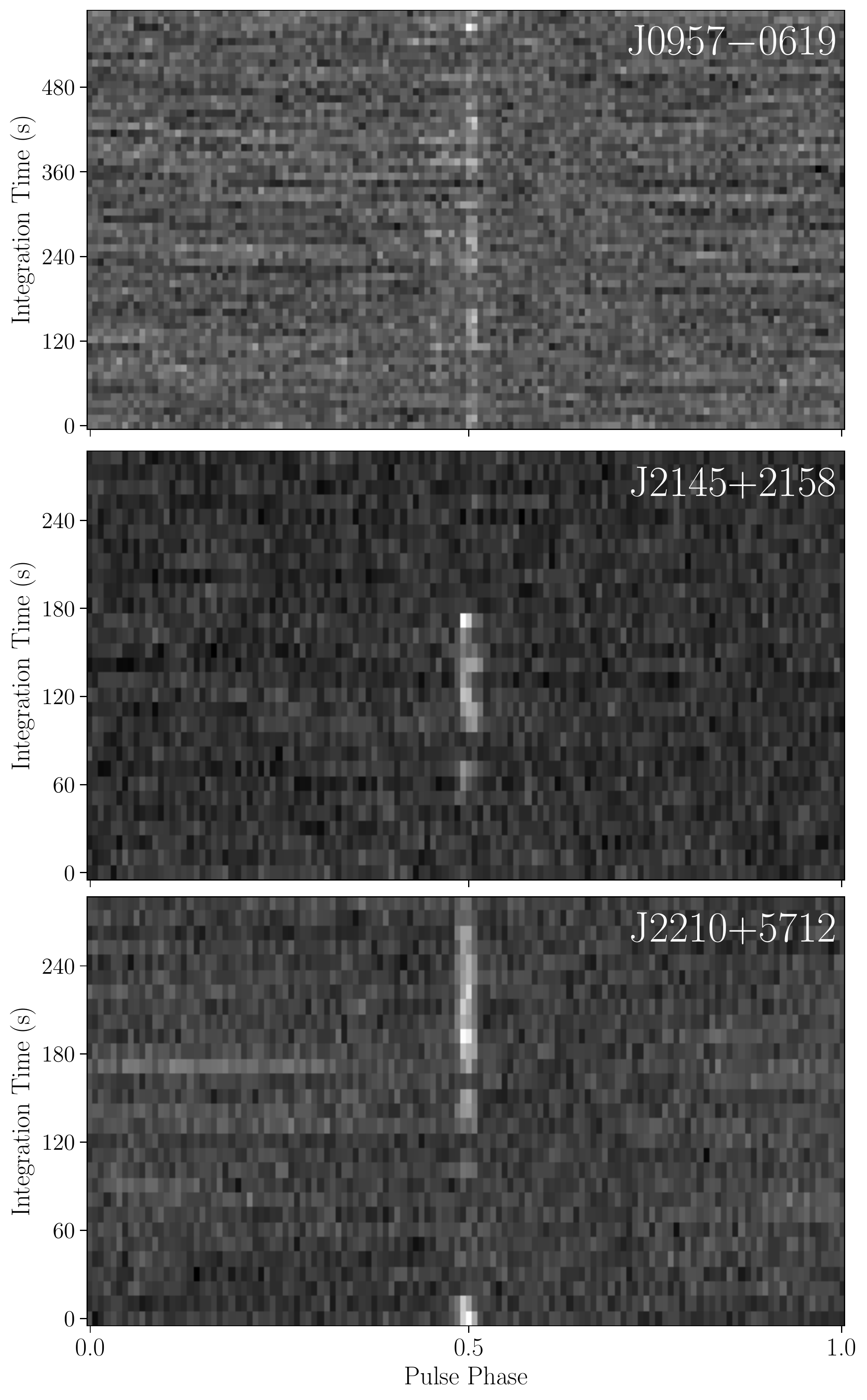}
\centering
\caption{Intensity of pulsar signal vs.~time and rotational phase for typical 820\,MHz observations of PSRs \psrf, \psrr, and \psrs, with greater intensity plotted as brighter and lower intensity plotted as darker. The nulling behavior of each pulsar is readily apparent, with the pulsar signal sometimes disappearing for well over a minute.}
\label{fig:nulling}
    \end{figure}

We noticed nulling behavior in each observation of PSRs \psrf, \psrr, and \psrs, with nulling periods lasting for a few seconds to a minute for \PSRf, and sometimes several minutes for PSRs \psrr and \psrs. Examples of typical observations are shown in Figure~\ref{fig:nulling}, clearly showing the evidence for nulling in each pulsar.

\begin{figure}
    \centering
    \includegraphics[width=0.48\textwidth]{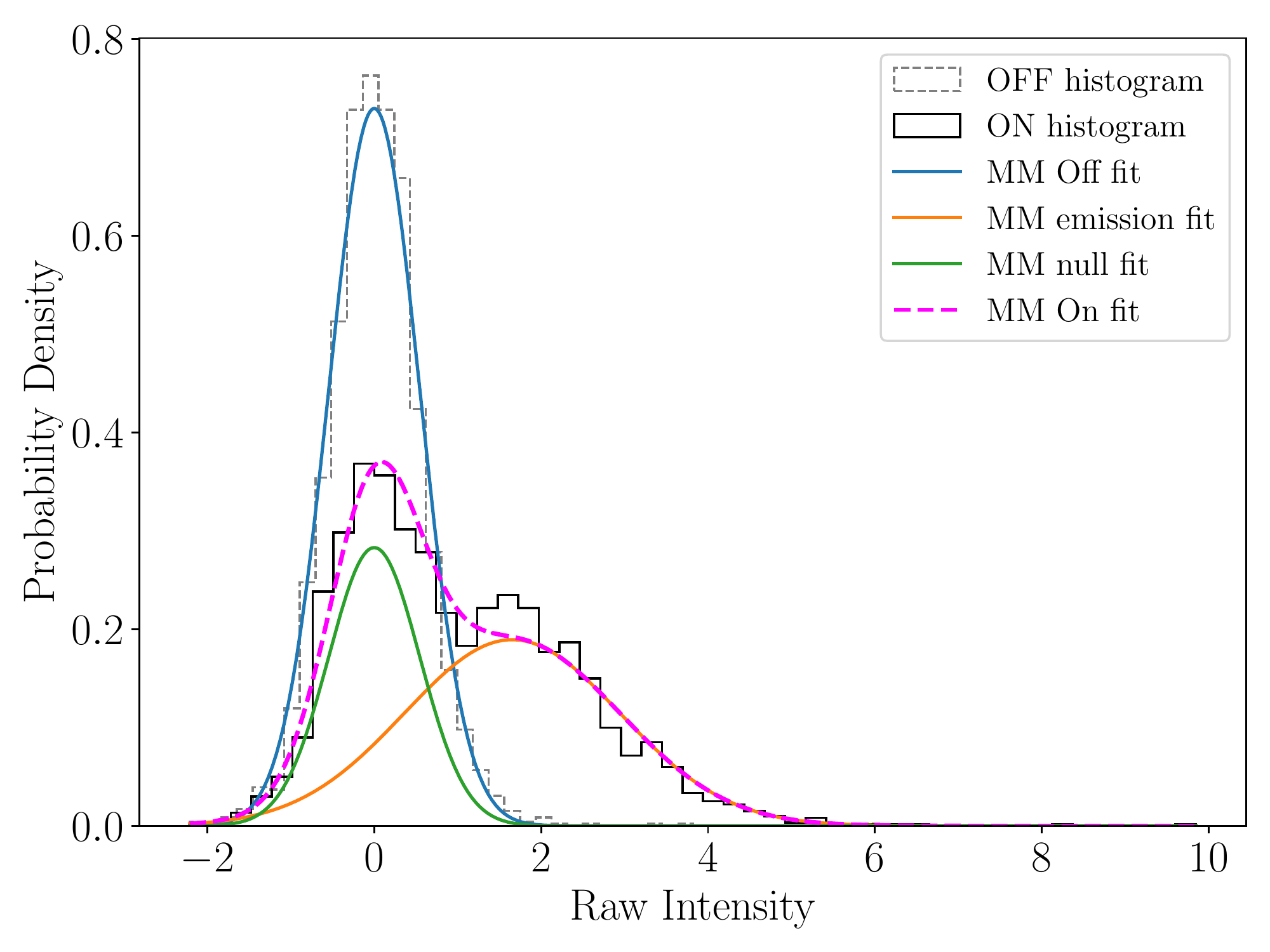}
    \caption{Histograms showing the intensity distributions of the pulsar emission and the background noise for \PSRs. 
    The dotted and solid histograms show the \textit{OFF} and \textit{ON} intensity distributions, respectively. 
    The presence of excess samples in the \textit{ON} distribution at the level consistent with background noise is evidence for nulling. 
    The blue line shows the Gaussian fit to the \textit{OFF} histogram.
    The fit to the \textit{ON} histogram, shown by the dashed magnenta line, is the sum of the ``emission'' and ``null'' components.
    The emission component (orange line) is a Gaussian centered above the background noise.
    The null component (green line) is a scaled-down version of the Gaussian noise; this scale factor gives the nulling fraction, which we measure to be $38.8\pm2.2$\% for this pulsar.}
    \label{fig:null_hist}
\end{figure}

We follow \citet{Kaplan2018, Akash2023} to estimate the nulling fraction using a Gaussian mixture model (GMM). 
A detailed description is provided in \citet{Akash2023}; we briefly describe it here.
We construct the \textit{ON} and \textit{OFF} histograms, which represent the distribution of intensities in a small window around the pulsar's emission phase and away from the pulsar's emission phase.
These are shown as the dotted and solid histograms in Figure~\ref{fig:null_hist}. 

The \textit{OFF} histogram is well-described by a Gaussian distribution, as expected for instrumental noise, assuming RFI has been sufficiently removed.
We model the \textit{ON} histogram as a Gaussian mixture of two components---a ``null'' component and an ``emission'' component. 
The intensities in the \textit{ON} distribution can be thought of as random draws from the emission component when the pulsed emission is observed. When a pulsar nulls, the intensities in the \textit{ON} distribution will be dominated by the background noise and hence this will be manifested as a scaled version of the \textit{OFF} distribution, which we call the ``null'' component.
The scale factor is called the nulling fraction (NF) of the pulsar. 

The component in the \textit{ON} histogram that is above the background noise represents the pulsar's emission and is called the ``emission'' component. 
By performing a simultaneous fit for both the \textit{ON} and \textit{OFF} histograms, we infer the NF (a detailed explanation of the fitting routine is provided in \citealt{Akash2023}).
We find the NF at 820\,MHz to be 59.0(2.3)\% for \PSRf, 67(2)\% for \PSRr and 38.8(2.2)\% for \PSRs.

\PSRf was discovered in the 350\,MHz Drift-scan survey as a RRAT, with a reported burst rate of 138(29)\,hr$^{-1}$ in a 10-minute GBT observation at 350\,MHz \citep{kkl+15}. 
As is apparent from the above analysis, this pulsar is not a RRAT at 820\,MHz, instead manifesting as a nulling pulsar. 
This is further evidence that RRATs may not represent a separate class of neutron star \citep{bb+10}.
Indeed, \citet{cbm+17} showed that RRAT emission likely represents the tail of the normal pulsar intensity distribution.

\subsection{Multiwavelength Counterparts}\label{sec:mwl}
Discussions of the $\gamma$-ray, X-ray, optical, and infrared counterparts of PSRs \psrc, \psre, and \psrl can be found in \citet{ksr+12}, \citet{kdk+13}, \citet{slr+14}, \citet{ska+16}, and \citet{ksk+18}.
The temperature of \PSRc's companion makes it either a very young WD or a subdwarf B star, and \PSRl's companion has properties which indicate it is a bloated proto-WD.

We searched for $\gamma$-ray counterparts of all new discoveries in the \textit{Fermi} LAT 12-year Source Catalog (4FGL-DR3), finding none.
We also used the Aladin server \footnote{\url{https://aladin.u-strasbg.fr/}} to perform a search for diffuse structures---for example, pulsar wind nebulae (PWN) and supernova remnants (SNRs) that may be associated with the 21 pulsars in this paper. 
We searched for symmetric diffuse structures expected for bow-shock PWNe (axisymmetric) and SNRs (radially symmetric) using all available imaging data across the electromagnetic spectrum for each pulsar and did find any candidates.

\begin{figure*}
    \centering
    \includegraphics[width=1\textwidth]{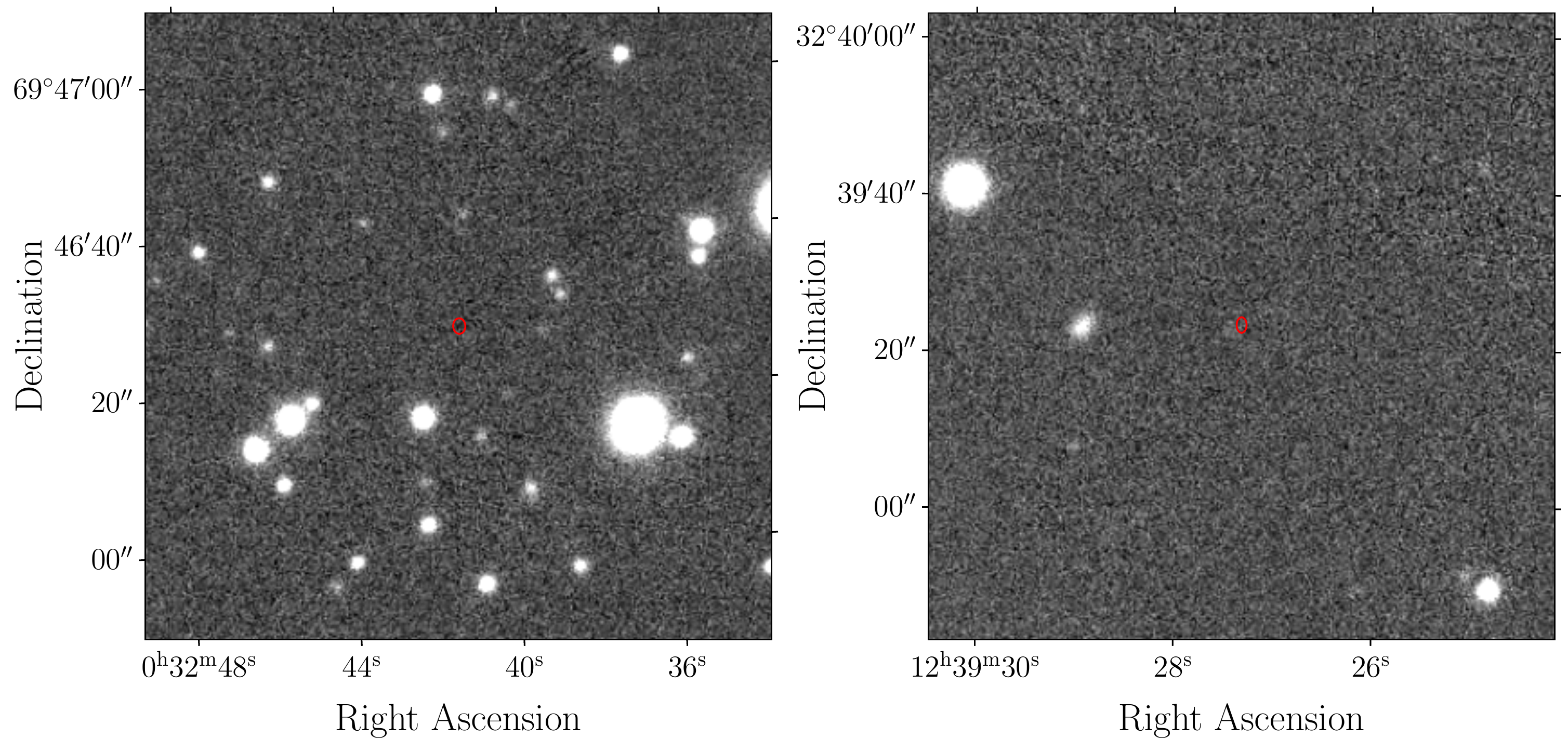}
    \caption{PanSTARRS1 $r$-band images, centered at the positions of \PSRa (left) and \PSRg (right). The red ellipses represent 1000\ $\times$ the uncertainties in the pulsar timing positions. No optical counterparts were found. There is a marginally-detected star near the position of \PSRg which also appears in the $i$-band image, but the distance between it and the pulsar is much larger than the frame-time uncertainty ($\sim 0.1\arcsec$).}
    \label{fig:ps1}
\end{figure*}

We checked images from the Panoramic Survey Telescope and Rapid Response System (PanSTARRS) $3\pi$ survey \citep{cmm+16} at the positions of the new binary pulsars \psra and \psrg, finding no optical counterparts coincident with the pulsars' timing positions. 
$r$-band images from the PanSTARRS1 (PS1) public science archive are shown in Figure \ref{fig:ps1}.
These undetected companion stars are most likely WDs. 

Following previous studies \citep{lsk+18,ksk+18,spp+23}, we use the PS1 nondetections to constrain the effective temperature $T_\mathrm{eff}$ and age of \PSRa's companion, assuming a CO-core WD.
The 5-$\sigma$ \texttt{grizy} magnitude limits are 23.3, 23.2, 23.1, 22.3, and 21.4, respectively \citep{cmm+16}.
Using a 3-D map of interstellar dust\footnote{\url{http://argonaut.skymaps.info/}} \citep{gsz+16}, we estimate the reddening to be 0.91 at a distance of 2.8\,kpc, corresponding to the DM-derived distance using the NE2001 Galactic electron density model.
We use that distance because it is larger than that predicted by the YMW16 model, and so it leads to a more conservative constraint.
We convert this to an extinction in each PS1 band using Table 6 in \citet{sfp+11}.

We compare our magnitude limits to cooling models\footnote{\url{https://www.astro.umontreal.ca/~bergeron/CoolingModels/}} for a 
0.5\,\Msun
WD \citep{bwd+11}, which is the companion mass assuming that $M_\mathrm{p} = 1.4$\,\Msun and $i = 60\arcdeg$.
The $r$-band limit provides the strictest constraints: $T_\mathrm{eff} < 45000 / 47500$\,K and age $> 1.70 / 1.74$\,Myr for hydrogen (DA) / helium (DB) atmospheres.

\begin{figure}
    \centering
    \includegraphics[width=0.5\textwidth]{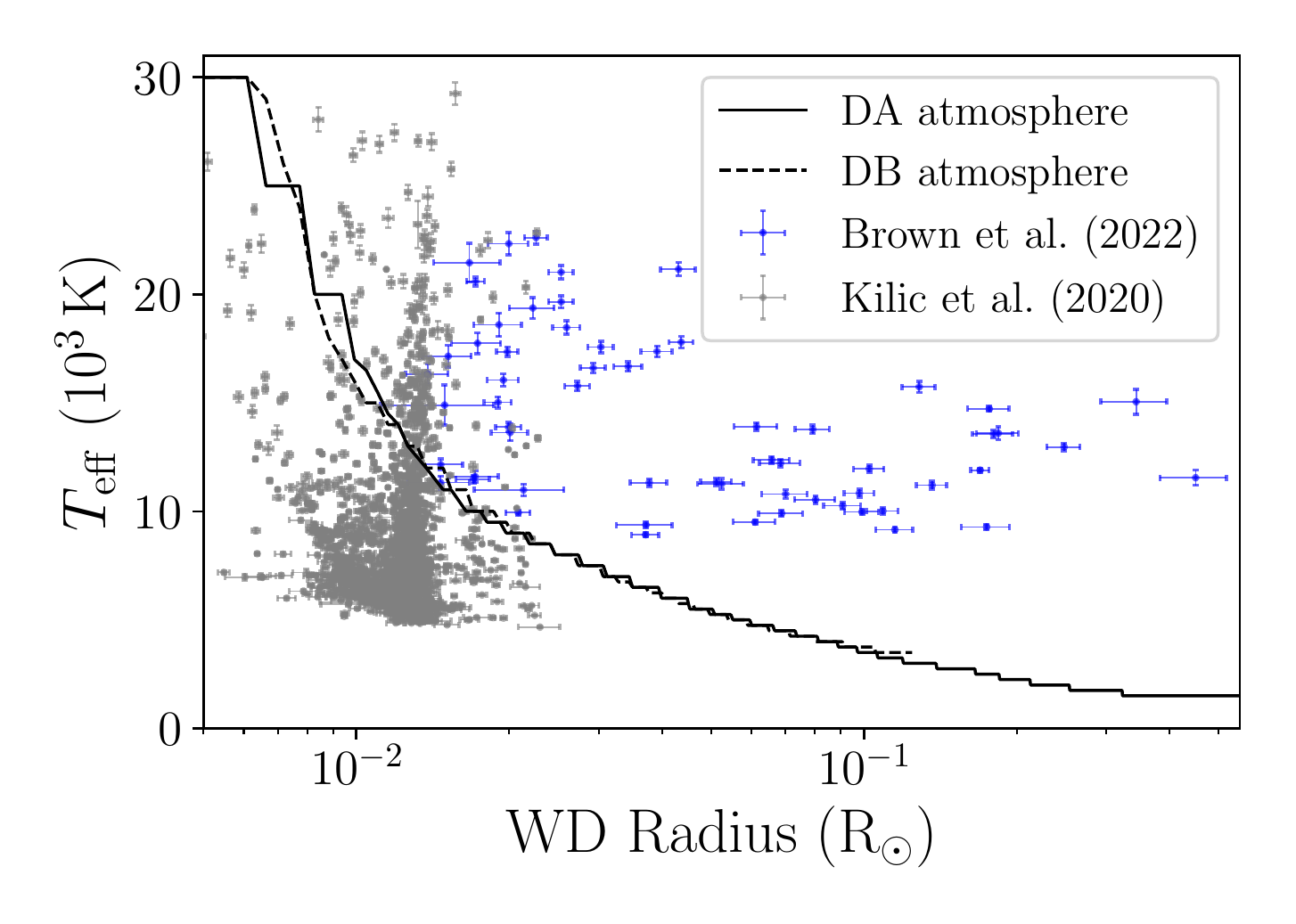}
    \caption{Limits on $T_\mathrm{eff}$ for \PSRg's $\sim$0.2\,\Msun companion, over a range of typical WD radii.
    The solid and dashed lines show the limit assuming DA or DB atmospheres, respectively.
    The gray points show WD radii and effective temperatures from WDs within 100\,pc in the SDSS footprint \citep{bkk+22}, and the blue points show values from the ELM Survey \citep{kbk+20}.
    }
    \label{fig:J1239teff}
\end{figure}

The lighter companion of \PSRg is probably an extremely low-mass \citep[ELM;][]{bkk+22} He-core WD.
Constraining the age is not simple, because hydrogen shell flashes and residual nuclear burning on the surfaces of He WDs lead to non-monotonic cooling \citep{apr+09}.
Taking the magnitudes given by the \citet{bwd+11} cooling models for a 0.2\,\Msun WD, we calculate new magnitudes by scaling the model radius to a range of possible WD radii.
We then compare these scaled magnitudes to our magnitude limits to infer the minimum $T_\mathrm{eff}$ which PS1 would detect for each radius, using the limit in the band that is most constraining---this varies between $g$, $r$, and $i$.
This gives us an upper limit curve constraining $T_\mathrm{eff}$, which is shown in Figure~\ref{fig:J1239teff}.

\vspace{1mm}
\section{Conclusion}\label{sec:conc}

We have presented pulse profiles, estimates of flux densities and spectral indices, and timing solutions for 20 pulsars discovered in the GBNCC pulsar survey, and one discovered in the GBT 350\,MHz Drift-scan pulsar survey. 
Three pulsars in our sample (\psrb, \psrh, and \psri) are DRPs, ejected from their binary systems by their former companion's supernova.
Contrary to what is expected of these systems, the transverse velocities we have measured for PSRs \psrh and \psri are not particularly high.

\PSRh has high timing precision for its spin period, and was observed by the NANOGrav PTA for $\sim$ 2 years at AO. 
We incorporated these observations into our timing analysis, allowing us to measure proper motion and parallax.
We also observed this pulsar using LWA1, and have shown that those observations hint at a chromatic DM, an effect caused by multipath scattering of the pulsar emission through small-scale density fluctuations in the ISM.

We presented new timing solutions for two new binary pulsars: \PSRa, a mildly recycled pulsar in a wide binary system; and \PSRg, an MSP orbiting a low-mass WD companion.
Their companion stars have masses consistent with the $P_\mathrm{b}$--$M_\mathrm{c}$ relation for WD companions, but are not seen in archival optical images, leading us to constrain the properties of the WDs.
We also presented updated timing solutions for \PSRc, another mildly recycled wide binary pulsar for which we weakly constrain $i$; \PSRe, a non-eclipsing black widow MSP which is observed by PTA experiments searching for low-frequency GWs; and \PSRl, an eclipsing binary MSP with a redback-mass companion.

We also analyzed three nulling pulsars using a Gaussian Mixture method.
One of these pulsars, \psrf, was discovered as a RRAT at 350\,MHz.
This is further evidence that RRATs may represent the tail of the intensity distribution of the general pulsar population.

The GBNCC pulsar survey has found 194 new pulsars in its coverage of the entire 350\,MHz GBT sky.
Follow-up timing of GBNCC discoveries continues at the Canadian Hydrogen Intensity Mapping Experiment (CHIME) telescope, thanks to collaboration between GBNCC and CHIME$-$Pulsar \citep{chime-pulsar}. 
Results from that effort will be presented in a future work.

\vspace{3.5mm}
\section*{Acknowledgements}\label{sec:ack}

Kevin Stovall performed the initial timing analysis of \PSRh and ran the LWA observations of that pulsar which are presented in this paper.
The Green Bank Observatory is a facility of the National Science Foundation (NSF) operated under cooperative agreement by Associated Universities, Inc.
The Arecibo Observatory is a facility of the NSF operated under cooperative agreement by the University of Central Florida and in alliance with Universidad Ana G.~Mendez, and Yang Enterprises, Inc.
Construction of the LWA has been supported by the Office of Naval Research under Contract N00014-07-C-0147 and by the AFOSR. Support for operations and continuing development of the LWA1 is provided by the Air Force Research Laboratory and the NSF under grants AST-1835400 and AGS1708855.
This paper is based in part on data obtained with the International LOFAR Telescope (ILT) under project code LC0\_002. LOFAR \citep{lofar} is the Low Frequency Array designed and constructed by ASTRON. It has observing, data processing, and data storage facilities in several countries, that are owned by various parties (each with their own funding sources), and that are collectively operated by the ILT foundation under a joint scientific policy. 
The ILT resources have benefitted from the following recent major funding sources: CNRS-INSU, Observatoire de Paris and Universit\'{e} d'Orl\'{e}ans, France; BMBF, MIWF-NRW, MPG, Germany; Science Foundation Ireland (SFI), Department of Business, Enterprise and Innovation (DBEI), Ireland; NWO, The Netherlands; The Science and Technology Facilities Council, UK.

W.F., M.A.M., D.L.K., J.K.S., M.E.D., T.D., S.M.R., and X.S.~are supported by the NANOGrav NSF Physics Frontiers Center award numbers 1430284 and 2020265. 
M.A.M.~and E.F.L.~are supported by NSF OIA-1458952 and NSF Award Number 2009425. 
T.D.~is supported by an NSF Astronomy and Astrophysics Grant (AAG) award number 2009468.
M.E.D.~acknowledges support from the Naval Research Laboratory by NASA under contract S-15633Y. 
V.M.K.~holds the Lorne Trottier Chair in Astrophysics \& Cosmology, a Distinguished James McGill Professorship, and receives support from an NSERC Discovery Grant (RGPIN 228738-13) and Gerhard Herzberg Award, from an R.~Howard Webster Foundation Fellowship from CIFAR, and from the Fonds de Recherche du Qu\'{e}bec: Nature et technologies, Centrede Recherche en Astrophysique du Qu\'{e}bec.
J.vL.~acknowledges funding from the European Research Council under the European Union’s Seventh Framework Programme (FP/2007-2013) / ERC Grant Agreement n.~617199 (``ALERT''), and from Vici research programme ``ARGO'' with project number 639.043.815, financed by the Netherlands Organisation for Scientific Research (NWO). 
S.M.R.~and I.H.S.~are CIFAR Fellows. 
Pulsar research at UBC is supported by an NSERC Discovery Grant and by the Canadian Institute for Advanced Research. 
E.P.~is supported by an H2020 ERC Consolidator Grant `MAGNESIA' under grant agreement No.~817661 and National Spanish grant PGC2018-095512-BI00. 
Z.P.~is a Dunlap Fellow. 
The work of X.S.~is partly supported by the George and Hannah Bolinger Memorial Fund in the College of Science at Oregon State University.
M.S.~acknowledges funding from the European Research Council (ERC) under the European Union’s Horizon 2020 research and innovation programme (grant agreement No.~694745). 

The Pan-STARRS1 Surveys (PS1) and the PS1 public science archive have been made possible through contributions by the Institute for Astronomy, the University of Hawaii, the Pan-STARRS Project Office, the Max-Planck Society and its participating institutes, the Max Planck Institute for Astronomy, Heidelberg and the Max Planck Institute for Extraterrestrial Physics, Garching, The Johns Hopkins University, Durham University, the University of Edinburgh, the Queen's University Belfast, the Harvard-Smithsonian Center for Astrophysics, the Las Cumbres Observatory Global Telescope Network Incorporated, the National Central University of Taiwan, the Space Telescope Science Institute, the National Aeronautics and Space Administration under Grant No.~NNX08AR22G issued through the Planetary Science Division of the NASA Science Mission Directorate, the NSF Grant No.~AST-1238877, the University of Maryland, Eotvos Lorand University (ELTE), the Los Alamos National Laboratory, and the Gordon and Betty Moore Foundation.

\facilities{GBT (GUPPI, VEGAS), AO (PUPPI), LOFAR, LWA1}

\software{
        \texttt{Astropy} \citep[\url{https://www.astropy.org/};][]{astropy:2022},
        \texttt{Matplotlib} \citep[\url{https://matplotlib.org/};][]{hjd+07},
        \texttt{Numpy} \citep[\url{https://numpy.org/};][]{numpy},
        \texttt{PINT} \citep{lrd+21},
        \texttt{PRESTO} \citep[\url{https://www.cv.nrao.edu/~sransom/presto/};][]{sem+02},
	\psrchive \citep[\url{https://psrchive.sourceforge.net/};][]{hsm+04, psrchive}, 
        \texttt{PyGDSM} \citep[\url{https://github.com/telegraphic/pygdsm};][]{pygsm},
	\texttt{PyGEDM} \citep[\url{https://github.com/FRBs/pygedm};][]{pygedm,yt+20},
        \texttt{Scipy} \citep[\url{https://scipy.org/};][]{scipy},
        \textsc{tempo} (\url{https://tempo.sourceforge.net/}),
        \tempotwo \citep[\url{https://www.atnf.csiro.au/research/pulsar/tempo2/};][]{hem+06, tempo2}
        }


\input{main.bbl}
\bibliographystyle{aasjournal}

\end{document}

%% file: duties-widths.tex
\begin{deluxetable*}{lcccccccccccc}
  \centering
  \tabletypesize{\footnotesize}
  \tablewidth{0pt}
  \tablecolumns{13}
  \tablecaption{Duty Cycles \& Equivalent Pulse Widths of GBNCC Pulsars}
  \tablehead{
    \colhead{}                                  &
    \multicolumn{2}{c}{149\,MHz}                &
    \multicolumn{2}{c}{350\,MHz}                &
    \multicolumn{2}{c}{430\,MHz}                &
    \multicolumn{2}{c}{820\,MHz}                &
    \multicolumn{2}{c}{\textit{1380}/1500\,MHz} &
    \multicolumn{2}{c}{2000\,MHz}              \\
    \colhead{PSR}                               &
    \colhead{$\delta$}                          &
    \colhead{$W_\mathrm{eq}$}                   &
    \colhead{$\delta$}                          &
    \colhead{$W_\mathrm{eq}$}                   &
    \colhead{$\delta$}                          &
    \colhead{$W_\mathrm{eq}$}                   &
    \colhead{$\delta$}                          &
    \colhead{$W_\mathrm{eq}$}                   &
    \colhead{$\delta$}                          &
    \colhead{$W_\mathrm{eq}$}                   &
    \colhead{$\delta$}                          &
    \colhead{$W_\mathrm{eq}$}                  \\
    \colhead{}                                  &
    \colhead{(\%)}                                  &
    \colhead{(ms)}                              &
    \colhead{(\%)}                                  &
    \colhead{(ms)}                              &
    \colhead{(\%)}                                  &
    \colhead{(ms)}                              &
    \colhead{(\%)}                                  &
    \colhead{(ms)}                              &
    \colhead{(\%)}                                  &
    \colhead{(ms)}                              &
    \colhead{(\%)}                                  &
    \colhead{(ms)}
  }
  \startdata
  J0032+6946 & \ldots & \ldots & \phn 3.84 & \dphn 1.42\phn & \ldots & \ldots     & \phn 2.23 & \dphn 0.82\phn & \ldots & \ldots & \ldots & \ldots \\
  J0141+6303 & \ldots & \ldots & \phn 5.07 & \dphn 2.37\phn & \ldots & \ldots     & \phn 2.78 & \dphn 1.3\dphn & \ldots & \ldots & \ldots & \ldots \\
  J0214+5222 & 5.81 & 1.428 & \phn 4.37 & \dphn 1.075 & \ldots & \ldots     & 21.32 & \dphn 5.24\phn & \ldots & \ldots & \ldots & \ldots \\
  J0415+6111 & \ldots & \ldots & \phn 4.4\phn & \phn 19.4\dphn & \ldots & \ldots     & \phn 4.34 & \phn 19.1\dphn & \ldots & \ldots & \ldots & \ldots \\
  J0636+5128 & \ldots & \ldots & \phn 4.76 & \dphn 0.136 & \ldots & \ldots     & \phn 4.97 & \dphn 0.143 & 5.58 & 0.16 & 5.42 & 0.156 \\
  J0957$-$0619 & \ldots & \ldots & \phn 0.51 & \dphn 8.8\dphn & \ldots & \ldots     & \phn 1.97 & \phn 34.0\dphn & \ldots & \ldots & \ldots & \ldots \\
  J1239+3239 & \ldots & \ldots & 24.63 & \dphn 1.158 & \ldots & \ldots     & 34.07 & \dphn 1.602 & \ldots & \ldots & \ldots & \ldots \\
  J1327+3423 & \ldots & \ldots & \phn 2.9\phn & \dphn 1.2\dphn & 3.0 & 1.25     & \phn 3.03 & \dphn 1.26\phn & 2.83 & 1.18 & \ldots & \ldots \\
  J1434+7257 & \ldots & \ldots & \phn 8.31 & \dphn 3.47\phn & \ldots & \ldots     & \phn 9.71 & \dphn 4.05\phn & 8.86 & 3.7 & \ldots & \ldots \\
  J1505$-$2524 & \ldots & \ldots & \phn 1.85 & \phn 18.5\dphn & \ldots & \ldots     & \phn 2.2\phn & \phn 21.9\dphn & \ldots & \ldots & \ldots & \ldots \\
  J1530$-$2114 & \ldots & \ldots & \phn 2.62 & \phn 13.2\dphn & \ldots & \ldots     & \phn 2.6\phn & \phn 13.1\dphn & \ldots & \ldots & \ldots & \ldots \\
  J1816+4510 & 54.1 & 1.728 & 26.24 & \dphn 0.838 & \ldots & \ldots     & 18.29 & \dphn 0.584 & 6.3 & 0.201 & \ldots & \ldots \\
  J1913+3732 & \ldots & \ldots & \phn 2.03 & \phn 17.3\dphn & \ldots & \ldots     & \phn 1.98 & \phn 16.8\dphn & \ldots & \ldots & \ldots & \ldots \\
  J1929+6630 & \ldots & \ldots & \phn 2.74 & \phn 22.1\dphn & \ldots & \ldots     & \phn 2.08 & \phn 16.8\dphn & \ldots & \ldots & \ldots & \ldots \\
  J1930+6205 & \ldots & \ldots & \phn 1.52 & \phn 22.2\dphn & \ldots & \ldots     & \phn 1.4\phn & \phn 20.3\dphn & \ldots & \ldots & \ldots & \ldots \\
  J2104+2830 & \ldots & \ldots & \phn 1.87 & \dphn 7.6\dphn & \ldots & \ldots     & \phn 1.62 & \dphn 6.6\dphn & \ldots & \ldots & \ldots & \ldots \\
  J2115+6702 & \ldots & \ldots & \phn 3.75 & \phn 20.7\dphn & \ldots & \ldots     & \phn 4.46 & \phn 24.6\dphn & \ldots & \ldots & \ldots & \ldots \\
  J2145+2158 & \ldots & \ldots & \phn 2.63 & \phn 37.4\dphn & \ldots & \ldots     & \phn 2.21 & \phn 31.3\dphn & \ldots & \ldots & \ldots & \ldots \\
  J2210+5712 & \ldots & \ldots & \phn 5.94 & 122.0\dphn & \ldots & \ldots     & \phn 2.18 & \phn 44.8\dphn & \ldots & \ldots & \ldots & \ldots \\
  J2326+6243 & \ldots & \ldots & \phn 5.05 & \phn 13.4\dphn & \ldots & \ldots     & \phn 3.59 & \dphn 9.5\dphn & \ldots & \ldots & \ldots & \ldots \\
  J2354$-$2250 & \ldots & \ldots & \phn 2.45 & \phn 13.7\dphn & \ldots & \ldots     & \phn 2.85 & \phn 15.9\dphn & \ldots & \ldots & \ldots & \ldots
  \enddata
  \tablecomments{
  Pulse duty cycles $\delta$ and widths $W_\mathrm{eq}$ of a boxcar pulse with equivalent height to the peak height of the pulse profile are listed for each pulsar in this analysis. The telescopes used were LOFAR (149\,MHz), GBT (350, 820, 1500, and 2000\,MHz), and AO (430 and 1380\,MHz). Measurements made at L-band (1380\,MHz with AO and 1500\,MHz with the GBT) are listed in the same column, with the 1380\,MHz measurements italicized.
  }
\end{deluxetable*}

%% file: LWA.tex
\begin{deluxetable}{ccccc}
  \centering
  \tabletypesize{\footnotesize}
  \tablewidth{0pt}
  \tablecolumns{5}
  \tablecaption{LWA1 Pulse Widths \& Flux Densities for \PSRh}
  \tablehead{
    \colhead{$\nu$}                    &
    \colhead{$t_\mathrm{int}$}         &
    \colhead{$\delta$}                 &
    \colhead{$W_\mathrm{eq}$}          &
    \colhead{$S_\nu$}                  \\
    \colhead{(MHz)}                    &
    \colhead{(s)}                      &
    \colhead{(\%)}                     &
    \colhead{(ms)}                     &
    \colhead{(mJy)}
    }
  \startdata
  35.1 & 14437 & 3.45 & 1.43 & 80(50) \\
  49.8 & 61508 & 3.41 & 1.42 & 2.1(1.3)$\times$10$^{2}$ \\
  64.5 & 57802 & 2.87 & 1.19 & 2.0(1.3)$\times$10$^{2}$ \\
  79.2 & 57802 & 2.47 & 1.02 & 1.8(1.1)$\times$10$^{2}$
  \enddata
  \tablecomments{
  At each center frequency $\nu$, the total integration time $t_\mathrm{int}$, pulse duty cycle $\delta$, width $W_\mathrm{eq}$ (that of a boxcar pulse with equivalent height to the peak height of the pulse profile), and estimated flux density $S_\nu$ are shown for \PSRh. 
  Values in parentheses are uncertainties, estimated as described in Section~\ref{sec:flux}.
  }
\end{deluxetable}

%% file: flux.tex
\begin{deluxetable*}{lrcccrcccccl}
  \centering
  \tabletypesize{\footnotesize}
  \tablewidth{0pt}
  \tablecolumns{12}
  \tablecaption{Flux Density \& Spectral Index Measurements of GBNCC Pulsars}
  \tablehead{
        \colhead{}                  &
	\multicolumn{2}{c}{350\,MHz} &
	\multicolumn{2}{c}{430\,MHz} &
	\multicolumn{2}{c}{820\,MHz} &
	\multicolumn{2}{c}{\textit{1380}/1500\,MHz} &
	\multicolumn{2}{c}{2000\,MHz} &
	\colhead{} \\
    \colhead{PSR}                      &
    \colhead{$t_\mathrm{int}$}            &
    \colhead{$S_\nu$}                  &
    \colhead{$t_\mathrm{int}$}            &
    \colhead{$S_\nu$}                  &
    \colhead{$t_\mathrm{int}$}            &
    \colhead{$S_\nu$}                  &
    \colhead{$t_\mathrm{int}$}            &
    \colhead{$S_\nu$}                  &
    \colhead{$t_\mathrm{int}$}            &
    \colhead{$S_\nu$}                  &
    \colhead{$\alpha$} \\
    \colhead{}                         &
    \colhead{(s)}                      &
    \colhead{(mJy)}                    &
    \colhead{(s)}                      &
    \colhead{(mJy)}                    &
    \colhead{(s)}                      &
    \colhead{(mJy)}                    &
    \colhead{(s)}                      &
    \colhead{(mJy)}                    &
    \colhead{(s)}                      &
    \colhead{(mJy)}                    &
    \colhead{}  
  }
  \startdata
  J0032+6946 & 13398 & 0.25(3)\phn & \ldots\phn & \ldots & 15567 & 0.19(5)\dphn & \ldots & \ldots & \ldots\phd & \ldots & $-$0.3(3) \\
J0141+6303 & 7352 & 0.57(6)\phn & \ldots\phn & \ldots & 11156 & 0.21(4)\dphn & \ldots & \ldots & \ldots\phd & \ldots & $-$1.2(3) \\
J0214+5222 & 4890 & 0.65(8)\phn & \ldots\phn & \ldots & 14395 & 0.37(6)\dphn & \ldots & \ldots & \ldots\phd & \ldots & $-$0.7(2) \\
J0415+6111 & 111 & 1.19(14) & \ldots\phn & \ldots & 5428 & 0.23(4)\dphn & \ldots & \ldots & \ldots\phd & \ldots & $-$1.9(3) \\
J0636+5128 & 14738 & 0.91(11) & \ldots\phn & \ldots & 28626 & 1.2(2)\tphn & 6770 & 0.17(4) & 5401 & 0.18(4) & $-$1.0(3) \\
J0957$-$0619 & 151 & 0.14(11) & \ldots\phn & \ldots & 10471 & 0.071(19) & \ldots & \ldots & \ldots\phd & \ldots & $-$1(1) \\
J1239+3239 & 12075 & 0.98(12) & \ldots\phn & \ldots & 13179 & 0.92(17)\phn & \ldots & \ldots & \ldots\phd & \ldots & $-$0.1(3) \\
J1327+3423 & 1421 & 1.5(3)\dphn & 31812 & 1.0(3) & 9373 & 0.66(15)\phn & \textit{30596} & \textit{0.086(19)} & \ldots\phd & \ldots & $-$1.9(2) \\
J1434+7257 & 7565 & 0.77(9)\phn & \ldots\phn & \ldots & 16370 & 0.25(5)\dphn & 594 & 0.10(2) & \ldots\phd & \ldots & $-$1.40(7) \\
J1505$-$2524 & 334 & 1.2(3)\dphn & \ldots\phn & \ldots & 6465 & 0.32(8)\dphn & \ldots & \ldots & \ldots\phd & \ldots & $-$1.5(4) \\
J1530$-$2114 & 1774 & 0.42(7)\phn & \ldots\phn & \ldots & 7945 & 0.13(3)\dphn & \ldots & \ldots & \ldots\phd & \ldots & $-$1.4(3) \\
J1816+4510 & 8737 & 1.29(15) & \ldots\phn & \ldots & 48013 & 0.21(4)\dphn & 32098 & 0.016(4) & \ldots\phd & \ldots & $-$2.9(4) \\
J1913+3732 & 546 & 3.5(7)\dphn & \ldots\phn & \ldots & 8691 & 2.1(6)\tphn & \ldots & \ldots & \ldots\phd & \ldots & $-$0.6(4) \\
J1929+6630 & 748 & 0.63(11) & \ldots\phn & \ldots & 4784 & 0.12(3)\dphn & \ldots & \ldots & \ldots\phd & \ldots & $-$2.0(4) \\
J1930+6205 & 605 & 0.43(12) & \ldots\phn & \ldots & 4784 & 0.08(3)\dphn & \ldots & \ldots & \ldots\phd & \ldots & $-$2.0(5) \\
J2104+2830 & 344 & 0.87(20) & \ldots\phn & \ldots & 5136 & 0.15(4)\dphn & \ldots & \ldots & \ldots\phd & \ldots & $-$2.1(4) \\
J2115+6702 & 111 & 0.65(9)\phn & \ldots\phn & \ldots & 5418 & 0.091(18) & \ldots & \ldots & \ldots\phd & \ldots & $-$2.3(3) \\
J2145+2158 & 30 & 1.6(3)\dphn & \ldots\phn & \ldots & 2386 & 0.16(4)\dphn & \ldots & \ldots & \ldots\phd & \ldots & $-$2.7(4) \\
J2210+5712 & 111 & 2.3(2)\dphn & \ldots\phn & \ldots & 5662 & 0.30(7)\dphn & \ldots & \ldots & \ldots\phd & \ldots & $-$2.4(3) \\
J2326+6243 & 111 & 1.48(15) & \ldots\phn & \ldots & 5428 & 0.75(14)\phn & \ldots & \ldots & \ldots\phd & \ldots & $-$0.8(2) \\
J2354$-$2250 & 1783 & 0.68(13) & \ldots\phn & \ldots & 5820 & 0.15(3)\dphn & \ldots & \ldots & \ldots\phd & \ldots & $-$1.8(4)
  \enddata
  \tablecomments{
  Total integration time ($t_\mathrm{int}$; not including periods of nulling or parts of observations removed due to RFI) used to generate profiles, and estimated flux densities ($S_\nu$), are shown for each pulsar in this analysis, for each observing band. 
  We also report calculated power-law spectral indices ($\alpha$).
  The telescopes used were the 
  GBT (350, 820, 1500, and 2000\,MHz), and AO (430 and 1380\,MHz). Measurements made with different telescopes at L-band (1380\,MHz with AO and 1500\,MHz with the GBT) are listed in the same column, with the 1380\,MHz measurements italicized.
  Values in parentheses are uncertainties, estimated as described in Section~\ref{sec:flux}.
  }
\end{deluxetable*}

%% file: rot-timing.tex
\begin{deluxetable*}{llrccccc}
  \tabletypesize{\footnotesize}
  \tablewidth{\textwidth}
  \tablecaption{Rotational and Timing Parameters of GBNCC Pulsars}
  \tablecolumns{8}
  \tablehead{
    \colhead{PSR} & \colhead{$\nu$} & 
    \colhead{\nudot} &
    \colhead{Epoch} & \colhead{Data Span} & 
    \colhead{$\delta t_\mathrm{RMS}$} & \colhead{$N_\mathrm{TOA}$} & 
    \colhead{EFAC} \\ 
    \colhead{} & \colhead{(Hz)} & 
    \colhead{(Hz\,s$^{-1}$)} &
    \colhead{(MJD)} & \colhead{(MJD)} & \colhead{(\us)} & \colhead{}}
  \startdata
  J0032+6946 & \phn27.171119572492(3) & $-$2.65006(3)$\times$10$^{-15}$ & 56736 & 55169--58303 & \dphn19.1 & 1430 & 1.08 \\
  J0141+6303 & \phn21.42232445491(4) & $-$7.65(2)$\times$10$^{-16}$ & 57431 & 57072--57789 & \dphn75.1 & \phn122 & 1.03 \\
  J0214+5222 & \phn40.691271761863(5) & $-$4.9002(7)$\times$10$^{-16}$ & 56974 & 55353--58594 & \dphn77.8 & \phn951 & 1.05 \\
  J0415+6111 & \dphn2.27174933348(6) & $-$2.8(4)$\times$10$^{-16}$ & 57234 & 57071--57397 & \phn609.4 & \dphn38 & 1.04 \\
  J0636+5128 & 348.55923172059(1) & $-$4.262(8)$\times$10$^{-16}$ & 56712 & 56027--57397 & \tphn1.9 & 1403 & 1.17 \\
  J0957$-$0619 & \dphn0.58014346794(2) & $-$5(1)$\times$10$^{-17}$ & 57220 & 57071--57369 & \phn767.0 & \dphn67 & 1.07 \\
  J1239+3239 & 212.71645129924(2) & $-$1.752(5)$\times$10$^{-16}$ & 57733 & 56054--59412 & \dphn21.5 & \phn283 & 1.13 \\
  J1327+3423 & \phn24.089008071282(2) & $-$7.514(3)$\times$10$^{-17}$ & 58067 & 57079--59055 & \tphn2.7 & 1575 & 1.18 \\
  J1434+7257 & \phn23.957175372381(1) & $-$3.1476(4)$\times$10$^{-16}$ & 56731 & 55196--58266 & \dphn31.4 & \phn351 & 1.11 \\
  J1505$-$2524 & \dphn1.000750227637(9) & $-$9.812(6)$\times$10$^{-16}$ & 57077 & 56754--57399 & \phn384.0 & \phn588 & 1.13 \\
  J1530$-$2114 & \dphn1.97887013684(2) & $-$1.817(1)$\times$10$^{-15}$ & 56994 & 56588--57399 & \phn443.1 & \dphn62 & 1.04 \\
  J1816+4510 & 313.17493532200(3) & $-$4.2246(5)$\times$10$^{-15}$ & 56945 & 55508--58382 & \tphn8.2 & \phn749 & 1.45 \\
  J1913+3732 & \dphn1.174979047088(6) & $-$1.9021(2)$\times$10$^{-15}$ & 56694 & 55988--57399 & \phn373.5 & \phn189 & 1.04 \\
  J1929+6630 & \dphn1.24066854069(5) & $-$1.079(6)$\times$10$^{-15}$ & 57136 & 56872--57399 & \phn289.2 & \dphn31 & 0.96 \\
  J1930+6205 & \dphn0.68675905720(8) & $-$7.85(5)$\times$10$^{-16}$ & 57027 & 56655--57399 & \phn828.6 & \dphn33 & 0.97 \\
  J2104+2830 & \dphn2.46469868711(3) & $-$5.930(9)$\times$10$^{-16}$ & 56743 & 56089--57397 & \phn293.6 & \dphn72 & 0.99 \\
  J2115+6702 & \dphn1.81119402046(5) & $-$5.5(4)$\times$10$^{-16}$ & 57235 & 57072--57397 & \phn693.7 & \dphn38 & 0.99 \\
  J2145+2158 & \dphn0.70472549203(2) & $-$1.105(2)$\times$10$^{-15}$ & 56928 & 56459--57397 & 1725.7 & \dphn60 & 1.48 \\
  J2210+5712 & \dphn0.48705587420(1) & $-$4.4(1)$\times$10$^{-16}$ & 57236 & 57072--57399 & \phn953.0 & \dphn74 & 1.11 \\
  J2326+6243 & \dphn3.75729497728(2) & $-$3.617(1)$\times$10$^{-14}$ & 57234 & 57072--57397 & \phn336.9 & \phn216 & 1.02 \\
  J2354$-$2250 & \dphn1.792046218541(1) & $-$1.287(1)$\times$10$^{-16}$ & 58207 & 56666--59748 & \phn286.1 & \dphn78 & 1.08 \\
  \enddata
  \tablecomments{For each pulsar in this analysis, measurements of spin frequency $\nu$ (at the listed reference epoch) and its derivative $\dot\nu$ are listed, with the $1$-$\sigma$ uncertainties on the last digit in parentheses. 
  Also listed are the dates spanned by the TOAs, timing residual rms $\delta t_\mathrm{RMS}$, number of TOAs $N_\mathrm{TOA}$, and EFAC, a scaling factor applied to TOA uncertainties which forces the reduced $\chi^2$ to equal unity. 
  All timing models use the DE440 solar system ephemeris and are referenced to the TT(BIPM2021) time standard.}
\end{deluxetable*}

%% file: pos-dm.tex
\begin{deluxetable*}{lllcrrcc}
  \tabletypesize{\footnotesize}
  \tablewidth{\textwidth}
  \tablecaption{Coordinates and DMs of GBNCC Pulsars}
  \tablecolumns{8}
  \tablehead{\colhead{PSR} & \multicolumn{3}{c}{Measured} & 
             \multicolumn{4}{c}{Derived} \\ 
             \colhead{} & 
             \colhead{$\alpha_\mathrm{J2000}$} &  
             \colhead{$\delta_\mathrm{J2000}$} & 
             \colhead{DM} &
             \colhead{$\ell$} & 
             \colhead{$b$} &
             \colhead{D$^\mathrm{NE2001}_\mathrm{DM}$} &
             \colhead{D$^\mathrm{YMW16}_\mathrm{DM}$} \\
             \multicolumn{3}{c}{} &
             \colhead{(\dmu)} &
             \colhead{($\arcdeg$)} &
             \colhead{($\arcdeg$)} &
             \colhead{(kpc)} &
             \colhead{(kpc)}
             }
  \startdata     
  J0032+6946 & $00^{\rm h}\, 32^{\rm m}\, 41\, \fs2477(3)$ & $+69\arcdeg\, 46\arcmin\, 28\, \farcs047(2)$ & 79.9988(2) & 121.30 & 6.96 & \phn2.8 & \phn2.3 \\
  J0214+5222 & $02^{\rm h}\, 14^{\rm m}\, 55\, \fs2746(2)$ & $+52\arcdeg\, 22\arcmin\, 40\, \farcs907(3)$ & 22.0371(3) & 135.63 & $-$8.42 & \phn1.0 & \phn1.2 \\
  J0141+6303 & $01^{\rm h}\, 41^{\rm m}\, 45\, \fs761(1)$ & $+63\arcdeg\, 03\arcmin\, 49\, \farcs445(9)$ & 272.762(2)\dphn & 128.60 & 0.75 & 44.3 & \phn8.8 \\
  J0415+6111 & $04^{\rm h}\, 15^{\rm m}\, 51\, \fs63(5)$ & $+61\arcdeg\, 11\arcmin\, 51\, \farcs8(3)$ & 70.8(1)\tphn & 145.15 & 7.49 & \phn2.3 & \phn1.8 \\
  J0636+5128 & $06^{\rm h}\, 36^{\rm m}\, 04\, \fs84705(3)$ & $+51\arcdeg\, 28\arcmin\, 59\, \farcs9658(6)$ & \phm{\Large{$^a$}}11.1075\tablenotemark{a}\phne & 163.91 & 18.64 & \phn0.5 & \phn0.2 \\
  J0957$-$0619 & $09^{\rm h}\, 57^{\rm m}\, 08\, \fs12(2)$ & $-06\arcdeg\, 19\arcmin\, 37\, \farcs5(9)$ & 27.3(1)\tphn & 244.83 & 36.20 & \phn1.2 & \phn2.5 \\
  J1239+3239 & $12^{\rm h}\, 39^{\rm m}\, 27\, \fs3140(1)$ & $+32\arcdeg\, 39\arcmin\, 23\, \farcs379(2)$ & 16.8590(1) & 147.36 & 83.89 & \phn1.5 & \phn2.2 \\
  J1327+3423 & $13^{\rm h}\, 27^{\rm m}\, 07\, \fs54861(3)$ & $+34\arcdeg\, 23\arcmin\, 37\, \farcs6777(8)$ & \phn\phm{\Large{$^a$}}4.1829\tablenotemark{a}\phne & 78.61 & 79.45 & \phn0.5 & \phn0.3 \\
  J1434+7257 & $14^{\rm h}\, 33^{\rm m}\, 59\, \fs7338(4)$ & $+72\arcdeg\, 57\arcmin\, 26\, \farcs495(1)$ & 12.6118(1) & 113.08 & 42.15 & \phn0.7 & \phn1.0 \\
  J1505$-$2524 & $15^{\rm h}\, 05^{\rm m}\, 22\, \fs529(3)$ & $-25\arcdeg\, 24\arcmin\, 50\, \farcs1(1)$ & 44.79(2)\dphn & 337.42 & 28.34 & \phn1.9 & \phn3.9 \\
  J1530$-$2114 & $15^{\rm h}\, 30^{\rm m}\, 43\, \fs00(4)$ & $-21\arcdeg\, 14\arcmin\, 21(2)\arcsec$ & 37.95(1)\dphn & 345.54 & 28.16 & \phn1.6 & \phn2.6 \\
  J1816+4510 & $18^{\rm h}\, 16^{\rm m}\, 35\, \fs9346(3)$ & $+45\arcdeg\, 10\arcmin\, 33\, \farcs855(3)$ & \phm{\Large{$^a$}}38.8881\tablenotemark{a}\phne & 72.83 & 24.74 & \phn2.4 & \phn4.4 \\
  J1913+3732 & $19^{\rm h}\, 13^{\rm m}\, 27\, \fs892(3)$ & $+37\arcdeg\, 32\arcmin\, 12\, \farcs35(3)$ & 72.29(2)\dphn & 69.10 & 12.13 & \phn4.2 & \phn7.6 \\
  J1929+6630 & $19^{\rm h}\, 29^{\rm m}\, 07\, \fs22(1)$ & $+66\arcdeg\, 30\arcmin\, 56\, \farcs0(1)$ & 59.74(8)\dphn & 98.01 & 21.11 & \phn4.3 & \phn8.2 \\
  J1930+6205 & $19^{\rm h}\, 30^{\rm m}\, 42\, \fs45(3)$ & $+62\arcdeg\, 05\arcmin\, 31\, \farcs7(2)$ & 67.5(3)\tphn & 93.66 & 19.39 & \phn5.7 & 10.7 \\
  J2104+2830 & $21^{\rm h}\, 04^{\rm m}\, 24\, \fs133(3)$ & $+28\arcdeg\, 30\arcmin\, 57\, \farcs58(4)$ & 62.16(5)\dphn & 74.18 & $-$12.18 & \phn3.7 & \phn5.7 \\
  J2115+6702 & $21^{\rm h}\, 15^{\rm m}\, 00\, \fs42(2)$ & $+67\arcdeg\, 02\arcmin\, 31\, \farcs8(4)$ & \phd\phm{\Large{$^a$}}54.5(1.3)\tablenotemark{b}\dphn & 104.05 & 12.49 & \phn2.7 & \phn2.9 \\
  J2145+2158 & $21^{\rm h}\, 45^{\rm m}\, 04\, \fs24(2)$ & $+21\arcdeg\, 58\arcmin\, 10\, \farcs9(3)$ & 44.3(1)\tphn & 75.72 & $-$23.37 & \phn2.8 & \phn5.4 \\
  J2210+5712 & $22^{\rm h}\, 10^{\rm m}\, 08\, \fs27(2)$ & $+57\arcdeg\, 12\arcmin\, 59\, \farcs8(4)$ & 192.9(2)\tphn\phn & 102.42 & 0.91 & \phn6.2 & \phn3.9 \\
  J2326+6243 & $23^{\rm h}\, 26^{\rm m}\, 41\, \fs492(7)$ & $+62\arcdeg\, 43\arcmin\, 22\, \farcs49(7)$ & 193.61(3)\tphn & 113.40 & 1.43 & \phn8.5 & \phn4.4 \\
  J2354$-$2250 & $23^{\rm h}\, 54^{\rm m}\, 17\, \fs688(3)$ & $-22\arcdeg\, 50\arcmin\, 05\, \farcs6(1)$ & 10.00(1)\dphn & 48.15 & $-$76.37 & \phn0.4 & \phn1.1
  \enddata
  \tablenotetext{a}{Fiducial DM value used in DMX model.}
  \tablenotetext{b}{Measured from the single observation with the highest S/N, held fixed in timing model.}
  \tablecomments{We report pulsar positions in Right Ascension and Declination referenced to the J2000 epoch ($\alpha_\mathrm{J2000}$ and $\delta_\mathrm{J2000}$, respectively), and dispersion measures (DMs) for each pulsar in this analysis. Values in parentheses are the $1$-$\sigma$ uncertainty in the last digit. We also present a set of parameters derived from measured positions and DMs: Galactic longitude $\ell$ and latitude $b$, and DM-derived distances D$_\mathrm{DM}$ using the NE2001 \citep{ne2001:2002} and YMW16 \citep{ymw+17} Galactic electron density models, as indicated. These distance estimates should be taken to have large fractional uncertainties, $\sim$30 -- 50\%~\citep{dgb+19}. In some cases, the reported precision in DM goes beyond expected month$-$year-timescale variability. We account for these changes using a DMX model (see Section~\ref{sec:dmx} and Figure~\ref{fig:DMX}) for PSRs~\psre, \psrh, and \psrl, and list here the reference DM used in that model. This treatment was not necessary for the other pulsars with such listed precision, so we simply note that the uncertainties listed for DMs are likely underestimated if they are $\lesssim$0.001\,\dmu.}
\end{deluxetable*}

%% file: derived.tex
\begin{deluxetable*}{llrllr}
  \tabletypesize{\footnotesize}
  \tablewidth{\textwidth}
  \tablecaption{Derived Common Properties of GBNCC Pulsars}
  \tablecolumns{6}
  \tablehead{
    \colhead{PSR} & \colhead{$P$} & \colhead{\pdot} &
    \colhead{$\tau_\mathrm{c}$} & \colhead{$B_\mathrm{surf}$} &
    \colhead{$\dot{E}$} \\
    \colhead{} & \colhead{(s)} & \colhead{(s\,s$^{-1}$)} &
    \colhead{(yr)} & \colhead{(Gauss)} & 
    \colhead{(\ergpersec)}}
  \startdata
  J0032+6946 & 0.036803783419083(3) & 3.58955(4)$\times$10$^{-18}$ & 1.6$\times$10$^{8}$ & 1.2$\times$10$^{10}$ & 2.8$\times$10$^{33}$ \\
  J0214+5222 & 0.024575294816350(3) & 2.9594(4)$\times$10$^{-19}$ & 1.3$\times$10$^{9}$ & 2.7$\times$10$^{9}$ & 7.9$\times$10$^{32}$ \\
  J0214+5222 & 0.024575294816349(3) & 2.9596(4)$\times$10$^{-19}$ & 1.3$\times$10$^{9}$ & 2.7$\times$10$^{9}$ & 7.9$\times$10$^{32}$ \\
  J0415+6111 & 0.44018941054(1) & 5.5(8)$\times$10$^{-17}$ & 1.3$\times$10$^{8}$ & 1.6$\times$10$^{11}$ & 2.5$\times$10$^{31}$ \\
  J0636+5128 & 0.00286895284644653(9) & 3.508(7)$\times$10$^{-21}$ & 1.3$\times$10$^{10}$ & 1.0$\times$10$^{8}$ & 5.9$\times$10$^{33}$ \\
  J0957$-$0619 & 1.72371155631(5) & 1.5(3)$\times$10$^{-16}$ & 1.8$\times$10$^{8}$ & 5.1$\times$10$^{11}$ & 1.2$\times$10$^{30}$ \\
  J1239+3239 & 0.0047010938453146(4) & 3.87(1)$\times$10$^{-21}$ & 1.9$\times$10$^{10}$ & 1.4$\times$10$^{8}$ & 1.5$\times$10$^{33}$ \\
  J1327+3423 & 0.041512709740513(3) & 1.2948(5)$\times$10$^{-19}$ & 5.1$\times$10$^{9}$ & 2.3$\times$10$^{9}$ & 7.1$\times$10$^{31}$ \\
  J1434+7257 & 0.041741147879765(2) & 5.4841(6)$\times$10$^{-19}$ & 1.2$\times$10$^{9}$ & 4.8$\times$10$^{9}$ & 3.0$\times$10$^{32}$ \\
  J1505$-$2524 & 0.999250334782(9) & 9.798(6)$\times$10$^{-16}$ & 1.6$\times$10$^{7}$ & 1.0$\times$10$^{12}$ & 3.9$\times$10$^{31}$ \\
  J1530$-$2114 & 0.505338870593(4) & 4.640(3)$\times$10$^{-16}$ & 1.7$\times$10$^{7}$ & 4.9$\times$10$^{11}$ & 1.4$\times$10$^{32}$ \\
  J1816+4510 & 0.0031931035571918(3) & 4.3073(5)$\times$10$^{-20}$ & 1.2$\times$10$^{9}$ & 3.8$\times$10$^{8}$ & 5.2$\times$10$^{34}$ \\
  J1913+3732 & 0.851079006454(4) & 1.3778(1)$\times$10$^{-15}$ & 9.8$\times$10$^{6}$ & 1.1$\times$10$^{12}$ & 8.8$\times$10$^{31}$ \\
  J1929+6630 & 0.80601705226(3) & 7.01(4)$\times$10$^{-16}$ & 1.8$\times$10$^{7}$ & 7.6$\times$10$^{11}$ & 5.3$\times$10$^{31}$ \\
  J1930+6205 & 1.4561147604(2) & 1.66(1)$\times$10$^{-15}$ & 1.4$\times$10$^{7}$ & 1.6$\times$10$^{12}$ & 2.1$\times$10$^{31}$ \\
  J2104+2830 & 0.405729108077(5) & 9.76(1)$\times$10$^{-17}$ & 6.6$\times$10$^{7}$ & 2.0$\times$10$^{11}$ & 5.8$\times$10$^{31}$ \\
  J2115+6702 & 0.55212196413(2) & 1.7(1)$\times$10$^{-16}$ & 5.3$\times$10$^{7}$ & 3.1$\times$10$^{11}$ & 3.9$\times$10$^{31}$ \\
  J2145+2158 & 1.41899223358(5) & 2.226(4)$\times$10$^{-15}$ & 1.0$\times$10$^{7}$ & 1.8$\times$10$^{12}$ & 3.1$\times$10$^{31}$ \\
  J2210+5712 & 2.05315252924(6) & 1.84(4)$\times$10$^{-15}$ & 1.8$\times$10$^{7}$ & 2.0$\times$10$^{12}$ & 8.4$\times$10$^{30}$ \\
  J2326+6243 & 0.266148919913(2) & 2.562(1)$\times$10$^{-15}$ & 1.6$\times$10$^{6}$ & 8.4$\times$10$^{11}$ & 5.4$\times$10$^{33}$ \\
  J2354$-$2250 & 0.5580213220250(4) & 4.008(3)$\times$10$^{-17}$ & 2.2$\times$10$^{8}$ & 1.5$\times$10$^{11}$ & 9.1$\times$10$^{30}$
  \enddata

  \tablecomments{We report properties derived from directly-measured quantities for each pulsar in this analysis: spin periods $P$, period derivatives \pdot, characteristic ages $\tau_\mathrm{c}$, inferred surface magnetic fields $B_\mathrm{surf}$, and spindown luminosities $\dot E$. These have not been corrected for apparent acceleration caused by kinematic effects. We calculate $\dot{E}$ and $B_\mathrm{surf}$ assuming a moment of inertia $I = 10^{45}$\,g\,cm$^2$; additionally, $B_\mathrm{surf}$ assumes a neutron star radius $R=10$\,km and $\alpha=90^{\circ}$ (angle between spin/magnetic axes). Calculating $\tau_\mathrm{c}$ relies on the assumption that spin-down is fully due to magnetic dipole radiation (braking index $n=3$) and that the initial spin period is negligible. Values in parentheses are the $1$-$\sigma$ uncertainty in the last digit.}
  \end{deluxetable*}

%% file: binaries.tex
\begin{deluxetable*}{clccccc}
  \tabletypesize{\footnotesize}
  \tablewidth{\textwidth}
  \tablecolumns{7}
  \tablecaption{Orbital Parameters for GBNCC Pulsars in Binary Systems}
  \tablehead{
      \multicolumn{2}{c}{Quantity} & 
      \colhead{\PSRa} &
      \colhead{\PSRc} &
      \colhead{\PSRe} &
      \colhead{\PSRg} &
      \colhead{\PSRl}
  }
  \startdata
  \cutinhead{Measured}
  $\nu_\mathrm{b}$ & (Hz) & 2.193632768(6)$\times10^{-8}$ & 2.260385778(9)$\times10^{-8}$ & 0.0001739119636(3) & 2.833032117(4)$\times10^{-6}$ & 3.2070609894(7)$\times10^{-5}$ \\
  $a\sin i/c$ & (s) & 178.674768(3) & 174.565762(5) & 0.0089858(1) & 2.371127(2) & 0.5954006(6) \\
  $T_\mathrm{asc}$ & (MJD) & 56399.134959(4) & 56339.115121(4) & 56711.9950666(4) & 57730.0800015(9) & 56945.0911546(2) \\
  $e\sin\omega$ & & \phm{$-$}0.00028554(2) & $-$0.00271145(5) & \phm{$-$}1(2)$\times10^{-5}$ & 5(2)$\times10^{-6}$ & \phm{$-$}8(2)$\times10^{-6}$ \\
  $e\cos\omega$ & & $-$0.00044865(3) & $-$0.00458640(6) & $-$1(2)$\times10^{-5}$ & 0(2)$\times10^{-6}$ & $-$1(2)$\times10^{-6}$ \\
  $\dot{\nu}_\mathrm{b}$ & (Hz\,s$^{-1}$) & \ldots & \ldots & $-$7.1(7)$\times10^{-20}$ & \ldots & \ldots \\
  $\ddot{\nu}_\mathrm{b}$ & (Hz\,s$^{-2}$) & \ldots & \ldots & $-$2.4(8)$\times10^{-27}$ & \ldots & \ldots \\
  $\dot{x}$ & (s\,s$^{-1}$) & \ldots & $-$2.1(6)$\times10^{-13}$ & \ldots & \ldots & \ldots \\
  \cutinhead{Derived}
  $P_\mathrm{b}$ & (days) & 527.621316(2) & 512.039767(2) & 0.0665513392(1) & 4.085401647(6) & 0.36089348198(8) \\
  $T_0$ & (MJD) & 56615.350(3) & 56638.6460(8) & 56712.02(2) & 57731.1(3) & 56945.19(1) \\
  $\omega$ & (\arcdeg) & 147.525(2) & 210.5913(6) & 145(106) & 89(22) & 97(14) \\
  $e$ & & 0.00053181(3) & 0.00532795(5) & 1(2)$\times10^{-5}$ & 5(2)$\times10^{-6}$ & 8(2)$\times10^{-6}$ \\
  $f_\mathrm{M}$ & (\Msun) & 2.20$\times10^{-2}$ & 2.18$\times10^{-2}$ & 1.76$\times10^{-7}$ & 8.58$\times10^{-4}$ & 1.74$\times10^{-3}$ \\
  $M_\mathrm{c}$ & (\Msun) & 0.417 & 0.416 & 0.007 & 0.126 & 0.162
  \enddata
  \tablecomments{We report binary parameters for each pulsar in this analysis with evidence of a binary companion. 
  These include measured orbital frequencies $\nu_\mathrm{b}$, semimajor axes projected along the line of sight $x=a\sin{i}/c$, times of ascending node $T_\mathrm{asc}$, and first and second Laplace-Lagrange parameters, $e_1 = e\sin\omega$ and $e_2 = e\cos\omega$. 
  For certain pulsars, we also report first and second derivatives of orbital frequency ($\dot{\nu}_\mathrm{b}$ and $\ddot{\nu}_\mathrm{b}$, respectively), and a time derivative of the projected semimajor axis, $\dot{x}$.
  Using the measured binary parameters, we derive the orbital period $P_\mathrm{b}$, time of periastron $T_0$, longitude of periastron $\omega$, eccentricity $e$, the binary mass function $f_\mathrm{M} = (M_\mathrm{c}\sin i)^3(M_\mathrm{p}+M_\mathrm{c})^{-2}$, and the minimum companion mass $M_\mathrm{c, min} = f_\mathrm{M}(M_\mathrm{p} = 1.4\,\mathrm{M}_\sun, i = 90\arcdeg)$.
  We used the ELL1 binary model for each pulsar, but with two different implementations: for PSRs \psra and \psrc, we implemented Equation~\ref{eq:roemer}, which expresses the Roemer delay up to third order in $e$, in \texttt{PINT}; for the other pulsars, we used the ELL1 binary model as implemented in \tempotwo, which only includes the first-order Roemer-delay terms.
  Values in parentheses are the $1$-$\sigma$ uncertainty in the last digit.
  }
\end{deluxetable*}

%% file: pms.tex
\begin{deluxetable*}{lcccccccccc}
\centering
  \tabletypesize{\footnotesize}
  \tablewidth{0pt}
  \tablecolumns{11}
  \tablecaption{Proper Motions and Kinematic Corrections for Five GBNCC Pulsars}
  \tablehead{
    \colhead{PSR}                                    &
    \colhead{$\mu_\alpha$}     &
    \colhead{$\mu_\delta$}       &
    \colhead{$D_\mathrm{DM}$} &
    \colhead{$v_\mathrm{t}$}            &
    \colhead{\pg}    &
    \colhead{\ps}    &
    \colhead{\pint}   &
    \colhead{$B_\mathrm{surf}$} &
    \colhead{$\tau_\mathrm{c}$} &
    \colhead{$\dot{E}$} \\
    \colhead{}                      &
    \colhead{(\pmu)} &
    \colhead{(\pmu)} &
    \colhead{(\kpc)} &
    \colhead{($\km\,\s^{-1}$)}      &
    \colhead{($10^{-21}$)}           &
    \colhead{($10^{-21}$)}           &
    \colhead{($10^{-19}$)}           &
    \colhead{($10^9\;\gauss$)} &
    \colhead{(Gyr)} &
    \colhead{($10^{33}\;\erg\,\s^{-1}$)}
  }
  \startdata
  J0214+5222 & 9(1) & 2(2) & 1.0(3) & 40(10) & $-$0.04 & 5.14 & 2.91 & 2.7 & 1.3 & 0.8 \\
   & & & 1.2(3) & 50(20) & $-$0.02 & 5.79 & 2.90 & 2.7 & 1.3 & 0.8 \\
  J0636+5128 & \phnm1.1(4) & $-$4.4(7) & 0.5(1) & 11(4)\phn & \phnm0.03 & 0.07 & 0.03 & 0.1 & 13.3 & 5.7 \\
   & & & \phn0.21(6) & 5(2) & \phnm0.01 & 0.03 & 0.03 & 0.1 & 13.1 & 5.8 \\
  J1327+3423 & $-$8.2(2) & \phnm4.3(4) & 0.5(1) & 21(6)\phn & $-$5.84 & 4.07 & 1.31 & 2.4 & 5.0 & 0.1 \\
   & & & 0.3(1) & 15(5)\phn & $-$4.92 & 2.94 & 1.31 & 2.4 & 5.0 & 0.1 \\
  J1434+7257 & $-$4.5(9) & $-$7.6(6) & 0.7(2) & 30(9)\phn & $-$4.96 & 5.63 & 5.48 & 4.8 & 1.2 & 0.3 \\
   & & & 1.0(3) & 40(10) & $-$5.87 & 7.69 & 5.47 & 4.8 & 1.2 & 0.3 \\
  J1816+4510 & 2(1) & $-$3(1)\phn\phd & 2.4(7) & 40(20) & $-$0.84 & 0.22 & 0.44 & 0.4 & 1.2 & 53.0 \\
   & & & 4(1)\phn\phd & 70(30) & $-$1.42 & 0.39 & 0.44 & 0.4 & 1.1 & 53.5
  \enddata
  \tablecomments{We report measured proper motions in right ascension $\mu_\alpha$ and declination $\mu_\delta$, with $1$-$\sigma$ uncertainties in the last digit given in parentheses. We list DM-based distance estimates, using the NE2001 (top) and YMW16 (bottom) Galactic electron density models, with $\approx$30\% uncertainty. Using these quantities, we calculate transverse velocities $v_t$, and corresponding corrections to $\dot P$, due to motion in the Galactic potential (\pg) and secular acceleration (Shklovskii effect; \ps). We then list the intrinsic value, $P_\mathrm{int} = \dot P - P_\mathrm{S} - P_\mathrm{G}$, and use that to re-calculate the surface magnetic field strength $B_\mathrm{surf}$, characteristic age $\tau_\mathrm{c}$, and spin-down luminosity $\dot{E}$.
  }
\end{deluxetable*}

%% file: misc.tex
\begin{deluxetable}{lcccc}
  \tabletypesize{\footnotesize}
  \tablewidth{\textwidth}
  \tablecaption{Additional Parameters for Two Pulsars}
  \tablecolumns{5}
  \tablehead{
  \colhead{PSR} & 
  \multicolumn{2}{c}{Measured} & 
  \multicolumn{2}{c}{Derived} \\
  \colhead{} &
  \colhead{FD1} &
  \colhead{$\varpi$} &
  \colhead{$\varpi_\mathrm{corr}$} &
  \colhead{$D_\varpi$} \\
  \colhead{} &
  \colhead{(\us)} &
  \colhead{(mas)} &
  \colhead{(mas)} &
  \colhead{(kpc)}
  }
  \startdata
  J0636+5128 & 3.0(1)$\times$10$^{-5}$ & \ldots & \ldots & \ldots \\
  J1327+3423 & 0.000164(2) & 4(1) & $1.1_{-0.5}^{+1.1}$ & $0.9_{-0.5}^{+0.8}$ \\
  \enddata
  \tablecomments{We report measurements of FD1, a profile frequency-dependency parameter, and a measurement of timing parallax ($\varpi$). Following \citet{vlm+10}, we give $\varpi_\mathrm{corr}$, corrected for Lutz-Kelker bias, and the corresponding parallax distance, $D_\varpi$. Values in parentheses are the $1$-$\sigma$ uncertainty in the last digit reported by \tempotwo. We give 68\% confidence intervals for the derived parameters.
  }
  \end{deluxetable}